\documentclass[a4paper,aps,prd,reprint,superscriptaddress]{revtex4-1}
\usepackage{amsmath}
\usepackage{amssymb}
\usepackage{upgreek}
\usepackage{epsfig,graphicx,textcomp,varioref}
\usepackage[usenames,dvipsnames]{xcolor}
\usepackage[colorlinks=true,linkcolor=Blue,urlcolor=Blue,citecolor=Blue]{hyperref}
\usepackage{subfig}
\usepackage{empheq}

\newcommand{\jcap}{jcap}
\newcommand\aap{A\&A}     

\begin{document}

\title{Wave Propagation in Modified Gravity}
\author{Jan \O.~Lindroos}
\email[]{jan.lindroos@ift.uib.no}
\affiliation{Institute for Physics and Technology, University of Bergen, N-5020 Bergen, Norway}
\affiliation{Institute of Theoretical Astrophysics, University of Oslo, N-0315 Oslo, Norway}

\author{Claudio Llinares}
\affiliation{Institute of Theoretical Astrophysics, University of Oslo, N-0315 Oslo, Norway}
\affiliation{Institute for Computational Cosmology, Department of Physics, Durham University, Durham DH1 3LE, U.K.}
\author{David F. Mota}
\affiliation{Institute of Theoretical Astrophysics, University of Oslo, N-0315 Oslo, Norway}


\date{\today}

\begin{abstract}
We investigate the propagation of scalar waves induced by matter sources in the context of scalar-tensor theories of gravity which include screening mechanisms for the scalar degree of freedom. The usual approach when studying these theories in the non-linear regime of cosmological perturbations is based on the assumption that scalar waves travel at the speed of light. Within General Relativity such approximation is good and leads to no loss of accuracy in the estimation of observables.  We find, however, that mass terms and non-linearities in the equations of motion lead to propagation and dispersion velocities significantly different from the speed of light. As the group velocity is the one associated to the propagation of signals, a reduction of its value has direct impact on the behaviour and dynamics of non linear structures within modified gravity theories with screening. For instance, the internal dynamics of galaxies and satellites submerged in large dark matter halos could be affected by the fact that the group velocity is smaller than the speed of light.  It is therefore important, within such framework, to take into account the fact that different part of a galaxy will see changes in the environment at different times.  A full non-static analysis may be necessary under those conditions.
\end{abstract}

\maketitle

\section{\label{sec:Introduction}Introduction}

Einstein's General Theory of Relativity is the foundation of our modern description of gravitational phenomena, ranging from stellar evolution and planetary dynamics to the evolution of the universe itself. However over the years it has become clear that in order for this theory to explain observations on galactic and cosmological scales it seems necessary to assume a universe dominated by dark energy and dark matter, neither of which can be explained within the standard model of particle physics. This has led to a resurgence of interest in modified theories of gravity, in the hope that they might help explain the observed universe \cite{Clifton:2011jh}. In this paper we look at one class of such models where an additional scalar degree of freedom is included in the gravitational sector, and the dynamics is governed by the action \eqref{action}, where the metric signature is chosen to be mostly positive. 

\begin{empheq}{align}	
S=&\int\sqrt{-g}\Big[\frac{M_{Pl}^2}{2}R-\frac{1}{2}\partial_{\mu}\phi\partial^{\mu}\phi-V(\phi)\Big]d^4x\notag\\
+&\int \mathcal{L}_{m}\big[\tilde{g}_{\mu\nu},\psi\big]d^4x\ \ \ ,\ \ \ \tilde{g}_{\mu\nu}=B(\phi)g_{\mu\nu}
	\label{action}
\end{empheq}
These models differ from quintessence models by the presence of the conformal coupling to matter $B(\phi)$, which gives rise to an effective potential dependent on the matter distribution. The conformal coupling, which is expected to be of gravitational strength, gives rise to an additional 'fifth force', whose existence has been heavily constrained by local tests on  deviations from the inverse square law, Casimir forces and violations of the Weak Equivalence Principle \cite{Mostepanenko:2008vb}. These constraints rule out most such models for natural values of the couplings, unless the fifth forces can be screened in high density regions where the constraints are most stringent. Models exhibiting such screening are often called screened modified gravity, where the chameleon \cite{Khoury:2003aq}, symmetron \cite{Hinterbichler:2010es} and galileon models \cite{2009PhRvD..79f4036N} are examples. In the chameleon model, the mass of the field depends on the local matter density in such a way that in high density regions, like on earth or in the solar system, it becomes large, leading to a Yukawa like suppression of the fifth force range. In low density regions on the other hand, the mass of the chameleon is small allowing long range fifth forces which can modify the predictions of GR on cosmological scales. In addition to the Yukawa screening, the chameleon model gives rise to additional screening through the so called thin-shell mechanism which restricts the fifth forces of objects like the sun and earth, to thin shells close to the surface \cite{Khoury:2010xi,Mota:2006fz}.

Usually the effects of the new scalar degree of freedom is studied in the static or quasi-static limit.  However, there is increasing interest in the possibility that scalar waves may yield non-negligible corrections to GR and have to be taken into account when making predictions for some of these theories.  Recently N-body simulations taking into account the full dynamical equations of motion for the symmetron model where performed \cite{Llinares:2013qbh, 2014PhRvD..89h4023L}, showing novel phenomena such as the creation of domain walls and their subsequent breakdown leading to energy release in the form of scalar waves.  Non-static effects were also studied in the linear and non-linear regime in different models in \cite{2014PhRvD..89b3521N, 2015JCAP...02..034B, zuma3,2015PhRvD..92f4005W}. The waves appearing in the symmetron model where shown to travel at the speed of light in accordance with the notion that the speed of sound equals the speed of light, $c_s=1$, for scalar fields with canonical kinetic terms \cite{Garriga:1999vw,Bertacca:2010ct,zuma4}. However from the field theoretic point of view, at least in the linear regime, waves in a massive scalar field originating from a localized source travel with the group velocity $c_g<1$ \cite{whitham2011linear}, contrary to the phase velocity of plane waves in the Fourier expansion, which travel travel at phase velocities larger than that of light, $c_p>1$. A similar thing happens when light travels through dispersive media, where the phase velocity can be larger than the speed of light in vacuum, but where the group velocity and the measured speed of light is smaller than the vacuum value. This stems from the fact that waves having a local origin are not described by a single plane wave mode, but rather by a distribution of modes or a wave packet, which collectively travel at the group velocity. These differences between the speed of propagation of light and the gravitational degrees of freedom may lead to observational consequences.  See for instance \cite{2015PhRvD..92h4061S} for some of the consequences of having different horizons for light and gravity.

The concept of group velocity is seldom mentioned in the literature on scalar field cosmology, and there seems to be some confusion as to the meaning of the different concepts of velocity (sound speed $c_s$, group velocity $c_g$ and phase velocity $c_p$). In this paper we review the relation between these concepts in the context of linear scalar waves, and investigate how these results carry over into the non-linear regime and curved spacetime using numerics. As an explicit example we look at the chameleon model, where we throughout assume that the non-minimal coupling in equation~\eqref{action} can be linearized, $\tilde{g}_{\mu\nu}\approx g_{\mu\nu}+\delta g_{\mu\nu}$ with $\delta g_{\mu\nu}<<1$. This allows us to linearize the matter coupling

\begin{empheq}{equation}
	\mathcal{L}_{m}(\tilde{g}_{\mu\nu},\psi)\approx\mathcal{L}_{m}(g_{\mu\nu},\psi)+\frac{\sqrt{-g}}{2}\delta B(\phi)T^{m}
	\label{linear_mass}
\end{empheq}
where $T^m$ is the trace of the Einstein frame stress-energy tensor $T_{m}^{\mu\nu}$ in the absence of scalar couplings
\begin{equation}
T^m=T_{m}^{\mu\nu}g_{\mu\nu}\ \ \ ,\ \ \ T_{m}^{\mu\nu}=\frac{2}{\sqrt{-g}}\frac{\partial \mathcal{L}_m(g_{\mu\nu},\psi)}{\partial g_{\mu\nu}}
\end{equation}
For the conformal coupling $B(\phi)$ and potential $V(\phi)$ we follow~\cite{Khoury:2003aq}:

\begin{empheq}{equation}
\label{ChamCoup}
V(\phi)=M^{4}\left(\frac{M}{\phi}\right)^{n}\ \ ,\ \ B(\phi)=e^{2\beta\phi}\approx1+2\beta\phi
\end{empheq}
which gives the equation of motion for the scalar field

\begin{empheq}{equation}
\label{Eq:full_CEOM}
\nabla_{\mu}\nabla^{\mu}\phi=S(\phi,T^{m})=V(\phi)_{,\phi}-\frac{1}{2}\delta B(\phi)_{,\phi}T^{m}. 
\end{empheq}
Throughout this paper we focus on perturbations in the scalar field induced by a matter source rather than how matter perturbations are affected by the presence of additional gravitational degrees of freedom. For modifications to the evolution and dynamics of ordinary matter perturbations in modified theories of gravity, see for instance \cite{2014PhRvD..90d4010R, Roshan:2015gra, zuma1, zuma2, Roshan:2015tva}.

The paper is structured as follows: In section~\ref{sec:ScalarWaves:LinReg} we look at the propagation of scalar waves in the linear regime and Minkowski spacetime. We then continue in section~\ref{sec:ScalarWaves:NonLinReg} by looking at how this behaviour is modified when the non-linearities are taken into account. This is followed by a brief discussion in section~\ref{sec:SpeedofSound} on the apparent inconsistency between the group velocity, $c_g\leq1$, determining the propagation speed of the scalar waves, and the effective speed of sound $c_s=1$ found in the fluid approach used in perturbation theory. Finally we summarize and conclude in section~\ref{sec:Summary}. 

\section{\label{sec:ScalarWaves:LinReg}Scalar Waves: Linear Regime}

We start by considering the simplest possible scenario, where we assume a Minkowski background, with non-relativistic matter $T^m\approx-\rho_m$. The matter distribution can then be split into a background and source part, $\rho_m=\rho_0(x^{\mu})+\delta\rho(x^{\mu})$ where we allow the background to potentially vary in both space and time (e.g. as a time varying cosmological background, or the static background solution for spherical source). This can be used to consider linear perturbations over potentially non-linear backgrounds, although this is beyond the scope of this paper. We can then write the field in terms of a background solution $\phi_0$ and a perturbation $\varphi$, $\phi=\phi_0(x^{\mu})+\varphi(x^{\mu})$ which allows us to Taylor expand the source term $S(\phi,\rho_m)$ in equation \eqref{Eq:full_CEOM} 

\begin{empheq}{equation}
S\approx\underbrace{\beta\rho_0-\frac{nM^{n+4}}{\phi_0^{n+1}}}_{S(\phi_0,-\rho_0)}+\frac{n(n+1)M^{n+4}}{\phi_0^{n+2}}\varphi+\beta\delta\rho
\end{empheq}
Using this approximation, the equations of motion for the perturbations $\varphi$ becomes

\begin{empheq}{equation}
\label{Eq:pert_CEOM}
\Big(\partial_{\mu}\partial^{\mu}-m^2\Big)\varphi=\beta\delta\rho\ \ ,\ \ m^2=\frac{n(n+1)M^{n+4}}{\phi_0(x_{\mu})^{n+2}}\ \ \ ,
\end{empheq}
where $\phi_0(x^{\mu})$ is the solution to the full equations for the background density $\rho_0(x^{\mu})$, and the approximation is valid as long as $|\varphi(x^{\mu})|<<\phi_0(x^{\mu})$. In this paper we restrict ourselves to a fixed matter background with the background field $\phi_0$ situated at the minima of the background potential. This yields the standard massive Klein-Gordon equation with the background field $\phi_0$ given by

\begin{empheq}{align}
\phi_0&=\left(\frac{nM^{n+4}}{\beta\rho_0}\right)^{1/(n+1)}\\
m^2&=(n+1)\frac{\beta\rho_0}{\phi_0}
\end{empheq}
From the expression for $\phi_0$ we can now estimate what densities justifies the linearization of the conformal coupling

\begin{empheq}{equation}
\phi_0<<\frac{1}{2\beta}\ \ \ ,\ \ \ \rho_0>>2nM^4(n\beta M)^n
\end{empheq} 
If we assume natural values for the coupling parameters $\beta\sim1/M_{Pl},\ M\sim\rho_{\Lambda}^{1/4}$~\cite{Mota:2006fz}, with $\rho_{\Lambda}$ being the vacuum energy density, then the ratio $M/M_{Pl}\sim10^{-30}$ and the approximation is valid for matter densities $\rho_m>>10^{-30n}\rho_{\Lambda}$, which covers most conceivable situations, for $n>0$.

\subsection{\label{sec:LinearCausal}Causality}

First we note that even though the plane wave solutions of equation \eqref{Eq:pert_CEOM} can travel at speeds exceeding that of light, the propagation of a signal originating from a localized source is always causal. This can be seen by considering the effect of a source on the future field where we can write the solution in terms of the retarded Greens function $G_{ret}$ \cite{Bogoliubov}, defined in terms of the fundamental equation for the future field.

\begin{empheq}{gather}
\Big(\partial_{\mu}\partial^{\mu}-m^2\Big)G_{ret}(x^\mu,x'^\mu)=\delta^4(x^\mu-x'^\mu)\notag\\
\tau=t-t'\ \ \ ,\ \ \ \lambda=\tau^2-|\mathbf{x}-\mathbf{x}'|^2\notag\\
G_{ret}(x^{\mu},x'^{\mu})=-\frac{\theta(\tau)}{2\pi}\left(\delta(\lambda)-\frac{m\theta(\lambda)J_{1}(m\sqrt{\lambda})}{2\sqrt{\lambda}}\right)
\label{Eq:Greens}
\end{empheq}
Here $J_{\nu}(x)$ are the Bessel functions of the first kind and $\theta(x)$ is the Heaviside step-function $\lambda=0$ defines the boundary of the future light cone of the source and we see that outside the light cone, $\lambda<0$, the Greens function vanishes. The particular solution $\varphi$ for an arbitrary source $\delta\rho$ can now be constructed by exploiting the linearity of the equation, which allows us to write $\varphi$ as a weighted superposition of Greens solutions for different $\delta$-sources

\begin{empheq}{align}
\label{Eq:LinInt}
\varphi=\beta\int\delta\rho(x'^{\mu})G_{ret}(x^{\mu},x'^{\mu})d^4x'
\end{empheq} 
It follows from this form of the solution, that the field at a spacetime point $(t,x)$ is only affected by the part of the source inside the past light cone of the spacetime point. This ensures that the propagation of perturbations caused by a localized source is causal and the upper limit for the propagation velocity is the speed of light $c=1$. As an example we consider the case of a point particle of unit mass created at the origin at $t=0$, which we model as a delta-source $\delta\rho=\theta(t)\delta^3(\mathbf{x})$. where assume that the scalar field is initially unperturbed $\varphi(t=0,\mathbf{x})=0$. The integral form of the solution in terms of the distance from the source $r$ then simplifies to

\begin{empheq}{align}
\label{Eq:LinInt_particle}
\varphi(t,r)=-\frac{\beta}{4\pi r}\left(1-mr\int^{u_f}_0\frac{J_1(u)du}{\sqrt{u^2+m^2r^2}}\right)\ \ \ ,
\end{empheq}
where $u_f=m\sqrt{t^2-r^2}$. In the infinite future the integral simplifies for finite values of $r$ to the standard Yukawa solution for a static $\delta$-source, where the negative sign of the solution is a manifestation of the force between equal charges, given by the gradient of the field $F_{\varphi}\propto-\nabla_r\phi$.  

\begin{figure}[htb]
  \begin{center}
    \includegraphics[width=90mm]{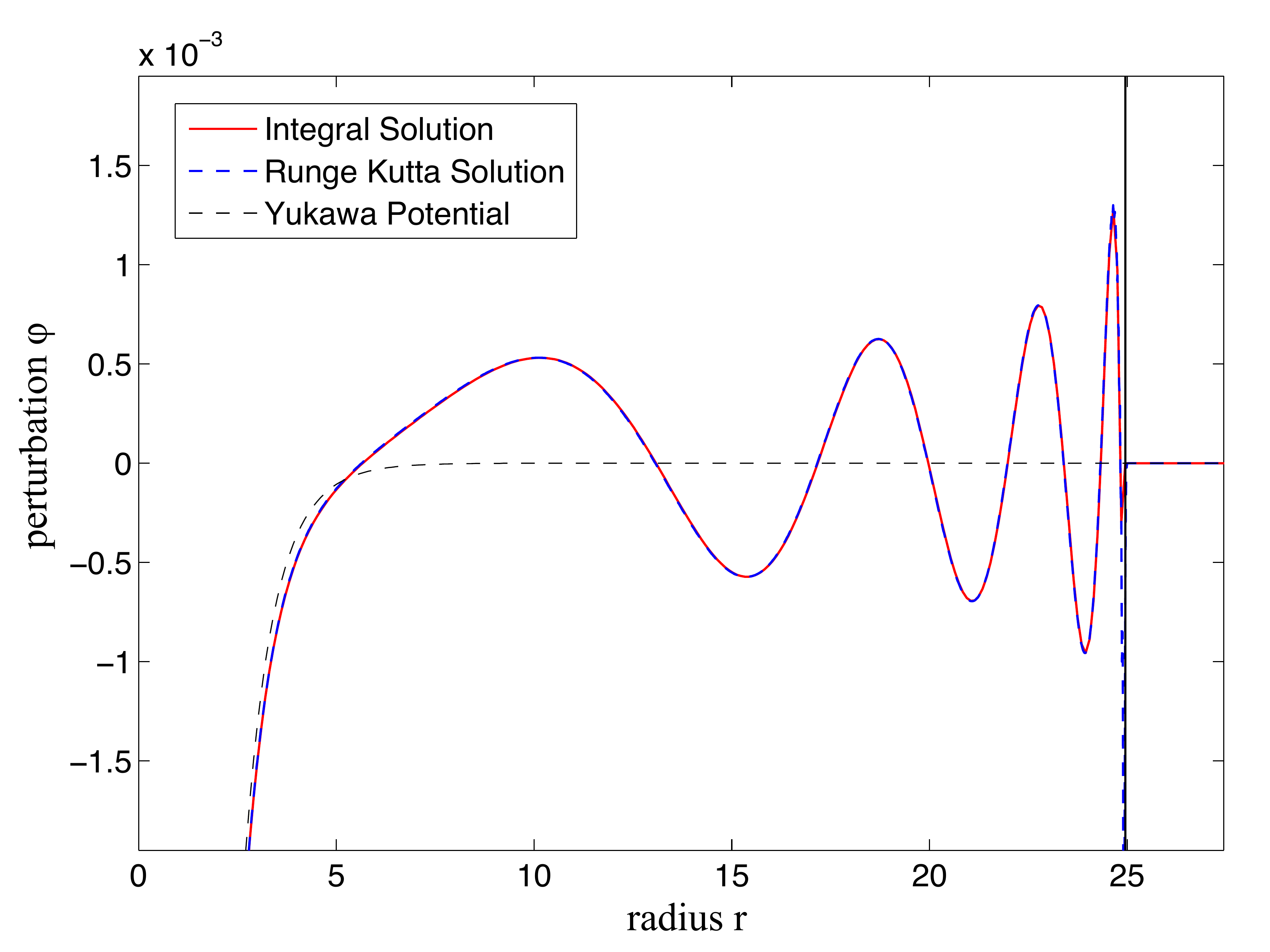}
    \caption{Perturbations $\varphi$ in the scalar field at $t=25$ induced by a static delta source created at the origin at $t=0$. The dashed blue and red lines corresponds to the solutions found by solving the differential \eqref{Eq:full_CEOM} and integral \eqref{Eq:Greens} equations respectively. The black vertical line indicates the light horizon, while the dashed black line corresponds to the static Yukawa solution. The length scale is arbitrary and related to the time scale by the speed of light $c=1$.}
    \label{fig:StaticDeltaSource}
  \end{center}              
\end{figure}

\begin{empheq}{align}
\label{Eq:StaticYukawa}
&\int^{\infty}_0\frac{J_1(u)}{\sqrt{u^2+m^2r^2}}=\frac{1-e^{-mr}}{mr}\notag\\
\Rightarrow\ \ \ &\lim_{t\rightarrow\infty}\varphi(t,r)=-\beta\frac{e^{-mr}}{4\pi r}
\end{empheq}
This is as expected since as $t\rightarrow\infty$, the motion created in the field due to the sudden appearance of the particle at $t=0$ has radiated away to infinity. In the finite case finding a closed form solution is difficult, but numerical solutions can be found either by solving the integral equation \eqref{Eq:LinInt} or the differential equation \eqref{Eq:full_CEOM}. Figure \ref{fig:StaticDeltaSource} shows the solutions at time $t=25$ for a scalar field with mass $m=1$, obtained using a 4th order Runge-Kutta solver for  eq. \eqref{Eq:full_CEOM} and an adaptive Gauss-Konrod quadrature integral solver \cite{Shampine2008131} for eq. \eqref{Eq:LinInt}. The light horizon is indicated by the vertical black line at $r=25$ after which no perturbations occur in accordance with the vanishing of the Greens function \eqref{Eq:Greens} beyond this point.      

The sudden introduction of a particle at $t=0$ creates oscillations in the scalar field which gradually subsides as the associated energy is radiated away. This leads to a wave pattern outside the source where the perturbations are large and narrow at the wavefront and gradually decay and widen in its wake as the central oscillations decay. As the oscillations decay the solution converges towards the Yukawa solutions \eqref{Eq:StaticYukawa} as predicted by the analytic solution. In the linear regime the perturbations in the field are proportional to the size of the matter perturbations and grows without bound as illustrated by the Yukawa solution. This is not necessarily true in the non-linear regime, and for chameleon models studied here, the self-coupling diverges as $\phi$ goes to zero, leading to a positive definite magnitude of the field. This results in a suppression of the scalar wave amplitude in the non-linear regime, due to smaller gradients in the field. The solutions found using the Runge-Kutta solver agrees well with the integral solutions, and serves as a sanity check for the differential solver that will be used for the remainder of the paper.
   
\subsection{\label{sec:WavePacket}Propagation of a Wave packet}

The concepts of phase and group velocity relates to the propagation of plane waves and wave packets respectively. To illustrate the concepts we consider the background equation ($\delta\rho=0$) for the perturbations $\varphi$ in Fourier space

\begin{empheq}{align}
\varphi(x^{\mu})&=\frac{1}{(2\pi)^4}\int \varphi(k^{\mu})e^{ik_{\mu}x^{\mu}}d^4k\label{eq:Fourier_x}\\
\varphi(k^{\mu})&=\int \varphi(x^{\mu})e^{-ik_{\mu}x^{\mu}}d^4x\label{eq:Fourier_k}\\
k^{\mu}&=\left( \omega,\mathbf{k}\right) 
\end{empheq}

which gives the Fourier space algebraic equation for any non-vanishing Fourier component $\varphi_k= \varphi(k^{\mu})$

\begin{empheq}{align}
&\left[\omega^2-(\mathbf{k}^2+m^2)\right]\varphi_{k}=0\\
&\Rightarrow\ \ \ \omega=\sqrt{\mathbf{k}^2+m^2}\label{Eq:disprel}
\end{empheq} 
The resulting relation between the frequency $\omega$ and the wave vector $\mathbf{k}$ is called the dispersion relation and tells us how the solutions to the equation propagate and disperse. The dispersion relation allows us to write the spacetime Fourier modes in terms of spatial ones \[\varphi_k=2\pi\varphi_{\mathbf{k}}\delta\left( \omega-\sqrt{\mathbf{k}^2+m^2}\right) \ \ ,\] 
and write equation \eqref{eq:Fourier_x} as

\begin{empheq}{equation}
\varphi(t,\mathbf{x})=\frac{1}{(2\pi)^3}\int \varphi_{\mathbf{k}}e^{-ik(c_pt-\hat{\mathbf{k}}\cdot\mathbf{x})}d^3\mathbf{k}
\label{eq:F_modes}
\end{empheq}
where the wave vector  is written in terms of its magnitude and directional unit vector $\mathbf{k}=k\mathbf{\hat{k}}$. Equation \eqref{eq:F_modes} tells us that the plane waves modes in the Fourier expansion travels with a velocity $c_p$, called the phase velocity.

\begin{empheq}{equation}
c_p=\frac{\omega}{k}=\sqrt{1+m^2/\mathbf{k}^2}>1
\end{empheq}
However the phase velocity describes the propagation of plane waves which are not localized, and hence different from the waves originating from a source. This is why the phase velocity can be greater than the speed of light, without breaking causality. A wave produced by a local source consists of a collection of plane waves, usually referred to as a wave packet, which never exceed the speed of light, as discussed in section \ref{sec:LinearCausal}. If the wave has a well defined wavelength $\lambda$, the propagation velocity can be estimated by considering solutions with $\varphi_{\mathbf{k}}$ sharply peaked around a wavenumber $k_0=1/\lambda$. For simplicity we consider propagation along one direction, in which case the frequency $\omega$ can then be Taylor expanded around $k_0$, allowing us to further simplify $\varphi(t,x)$ \cite{citeulike:486037}.

\begin{empheq}{align}
\omega(k)&\approx\omega(k_0)+\frac{d\omega(k_0)}{dk}\delta k=\omega_0+c_g(k-k_0)
\label{eq:freq_exp}\\
\varphi(t,x)&=\frac{e^{i[c_gk_0-\omega_0]t}}{2\pi}\int \varphi_{\mathbf{k}}e^{-ik(c_gt-x)}dk\notag\\
&=\varphi_0(x-c_gt)e^{i\theta_0t}\ \ \ ,\ \ \ \theta_0=c_gk_0-\omega_0\label{eq:wavepack_sol}
\end{empheq}
As can be seen from equation \eqref{eq:wavepack_sol}, the wave packet travels undistorted up to a global phase factor $\theta_0$ with a velocity $c_g$  as long as higher order terms in the frequency expansion are neglected. The velocity $c_g$ is what is called the group velocity and is given by

\begin{empheq}{equation}
c_g=\frac{d\omega}{dk}=\frac{1}{\sqrt{1+\frac{m^2}{k^2}}}<1
\label{eq:group_velocity}
\end{empheq}
Contrary to the phase velocity this is always smaller than the speed of light. To illustrate we consider a 1-dimensional oscillating Gaussian matter perturbation $\delta\rho$ oscillating with a frequency $\omega_m$.

\begin{figure*}[htbp!]
  	\subfloat[$m=0.01$]{\includegraphics[width=0.33\linewidth]{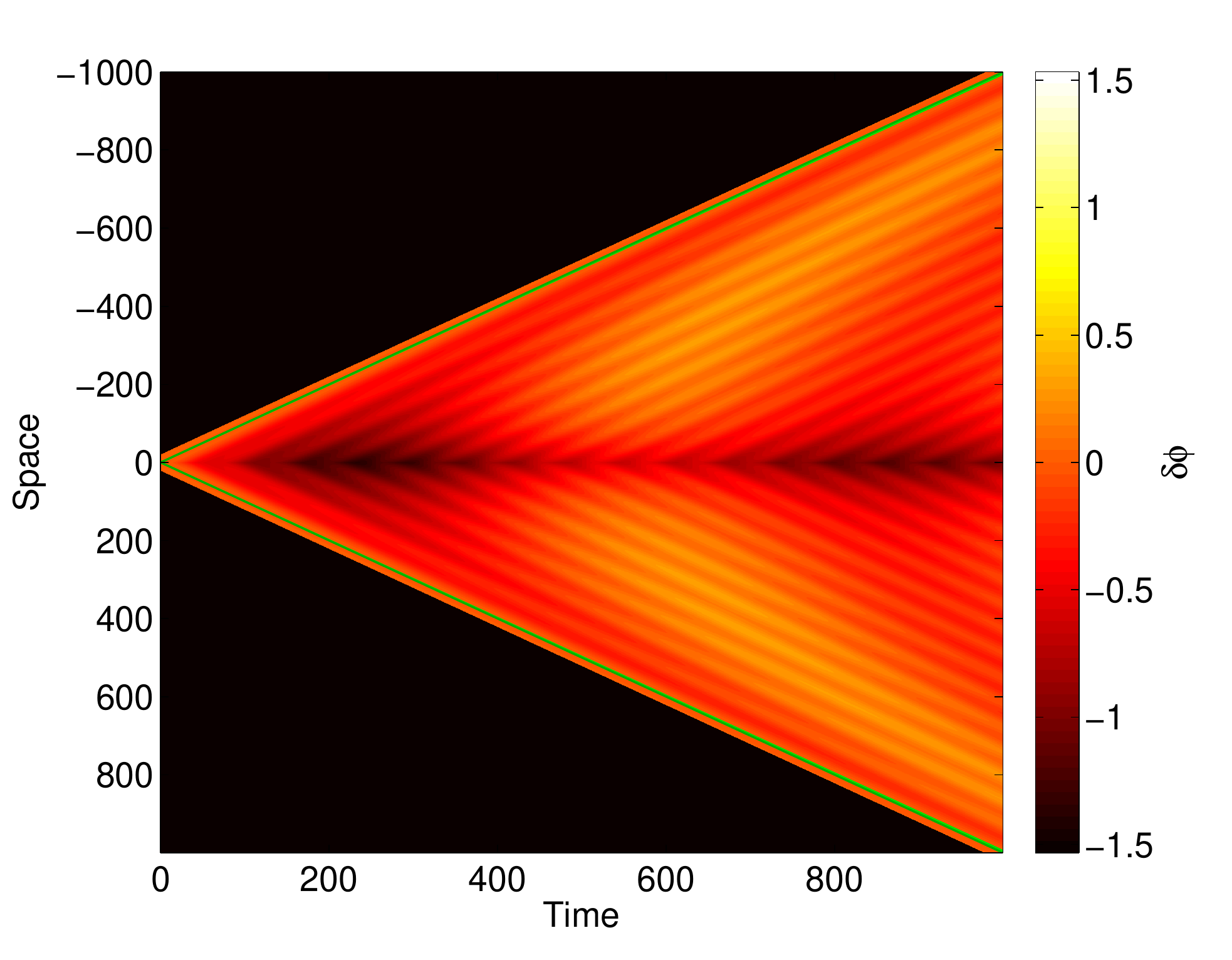}\label{fig:lc_m001_linear}}
    \subfloat[$m=0.1$]{\includegraphics[width=0.33\linewidth]{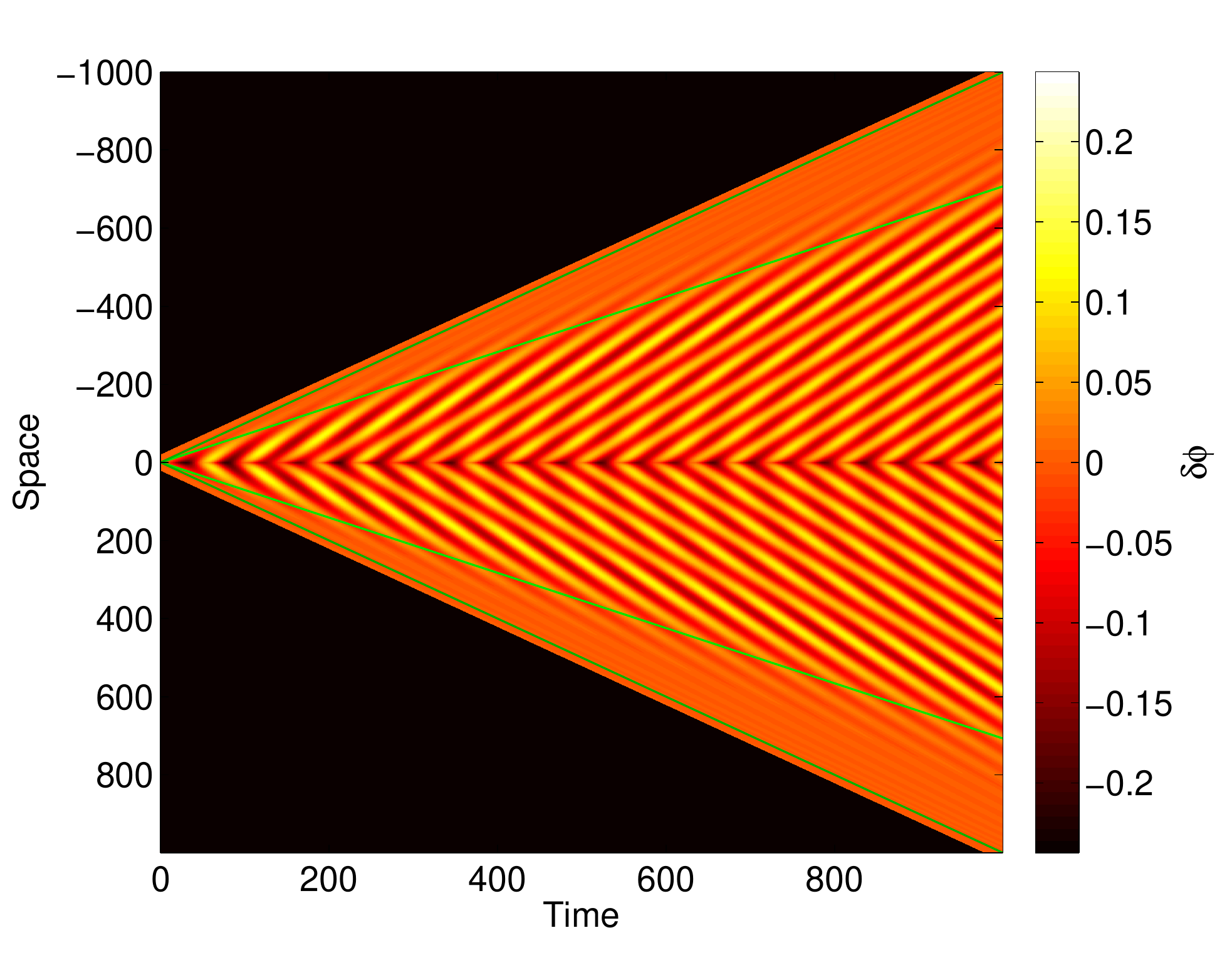}\label{fig:lc_m01_linear}}
    \subfloat[$m=1$]{\includegraphics[width=0.33\linewidth]{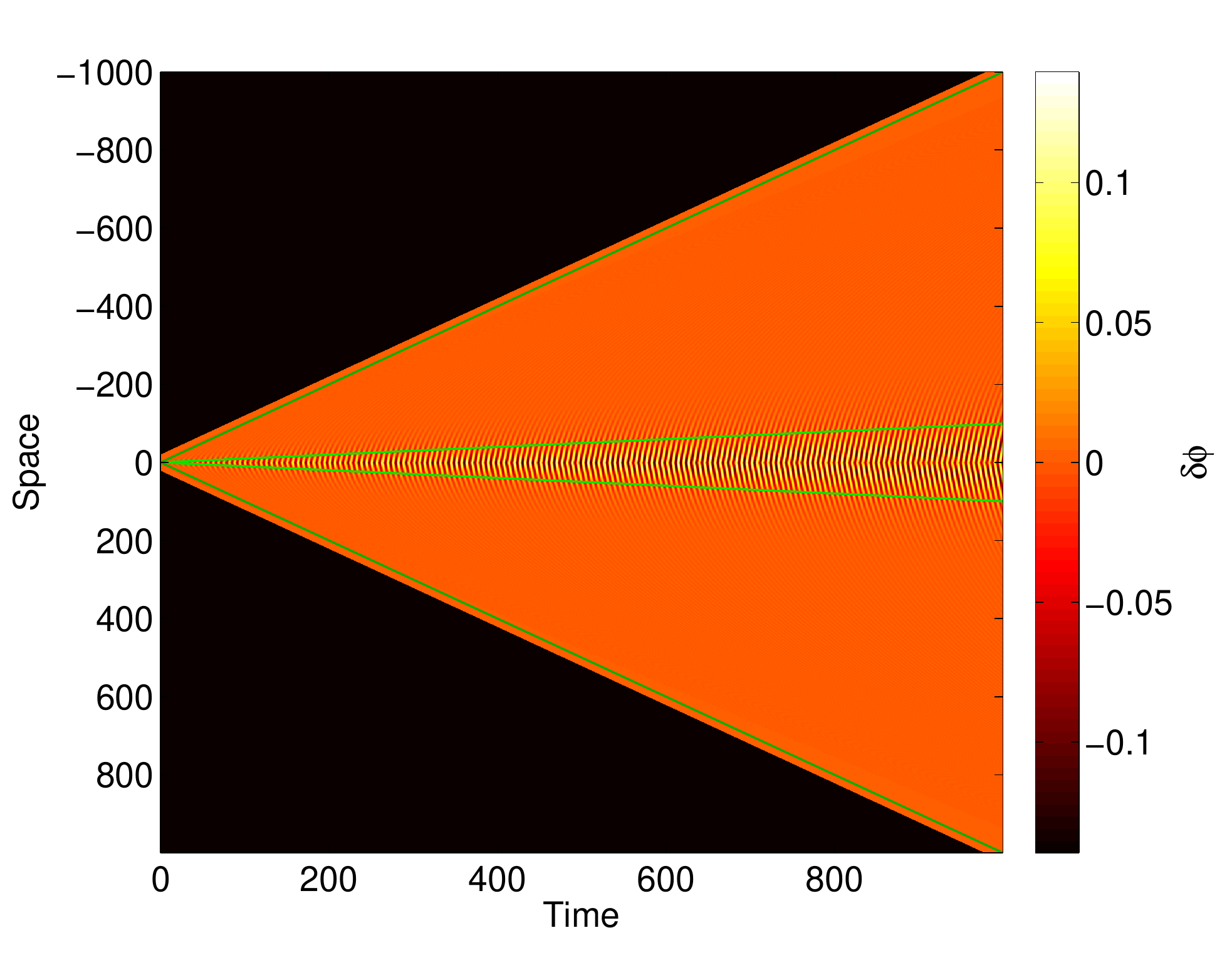}\label{fig:lc_m1_linear}}\\
    \subfloat[$m=0.1\ \ \ ,\ \ \ v_g=0.7c$]{\includegraphics[width=\linewidth]{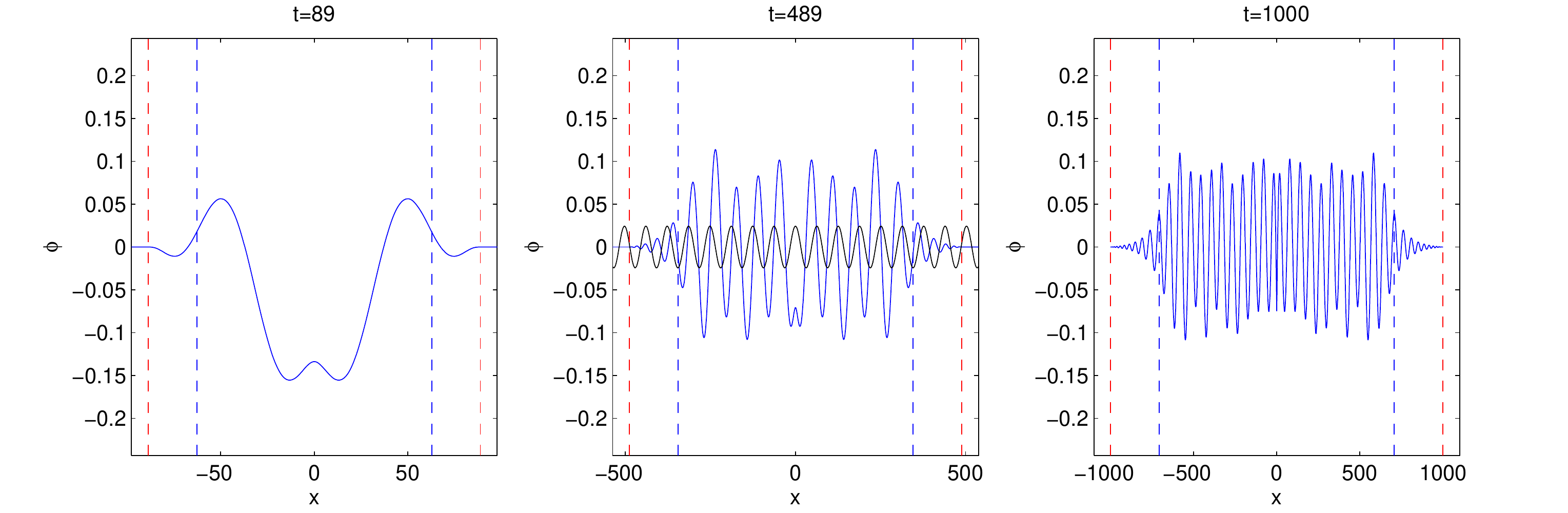}\label{fig:ts_m01_linear}}
    \caption{Plots \protect\subref{fig:lc_m001_linear}, \protect\subref{fig:lc_m01_linear}and \protect\subref{fig:lc_m1_linear} show the propagation of waves from a source for 3 different scalar masses, represented in terms of light cone plots. The lines indicate the light cone (dark green, wide angle) and the 'sound' cone (light green, small angle) indicating the front of the main signal as determined by the group velocity. The unperturbed regions are coloured in black for clarity. Plot \protect\subref{fig:ts_m01_linear} shows the corresponding scalar field profile for $m=0.1$ at three different times, where the plane wave $\sin(k_0x)$ is included for comparison in the middle plot.  The blue and red vertical dashed lines show the light horizon and the horizon given by the group velocity.}
    \label{lightcone_fig}            
\end{figure*}

\begin{empheq}{align}
&\delta\rho=\frac{1}{2}\delta\rho_0(x)(1-\cos(\omega_m t))\\ &\delta\rho_0(x)=Me^{-\frac{x^2}{2\sigma^2}}\ \ \ ,\ \ \ \sigma=0.1
\label{eq:osc_matter}
\end{empheq}
The particular solution to the one dimensional equation in Fourier space is then given by

\begin{equation}
	\varphi_{\mathbf{k}}=-\frac{\beta}{2}\delta\rho_{k0}\left[\frac{\cos\left(\omega_m t\right)}{\omega_m^2-\omega_{k}^2}+\omega_m^{-1}\right]
	\label{Eq:fourier_sol}
\end{equation}

where $\omega_{k}$ is th frequency associated with $\mathbf{k}$ through the dispersion relation \eqref{Eq:disprel}. By choosing an oscillation frequency $\omega_m=\sqrt{k_0^2+m^2}$, the source induces perturbations in $\varphi$ dominated by $k_0$. Figure \ref{lightcone_fig} shows the evolution of the scalar field in a similar fashion as in \cite{Llinares:2013qbh},  for three different scalar masses $m=[0.01,0.1,1]$, $\beta=1$ and $k_0=M=0.1$, in addition to the profile for $m=0.1$ at three different times.  The unperturbed regions in the light cone plots are coloured in black for clarity.
 
The solutions show the main component of the wave traveling at the group velocity, and a rapid decay of the solution in front. The small high frequency oscillations in front of the main signal are called precursors and is a well known phenomena from light propagation in dispersive media  \cite{ANDP:ANDP19143491003,ANDP:ANDP19143491002}. These precursors are generally much smaller than the main signal. The same behavior is also observed for a spherically symmetric source except for an additional $1/r$ decay of the amplitude as we will see when we solve the full equations in section~\ref{sec:ScalarWaves:NonLinReg}.

Including higher order terms in the frequency expansion in equation \eqref{eq:freq_exp} leads to dispersion and amplitude evolution.  In particular we can get an approximate expression for the dispersion rate (i.e how fast the wave packet spreads) by including the second order term in the expansion. Let us consider for example, a spherically symmetric perturbation at rest with an initial width $\sigma(t=0)=\sigma_{0}$:
\begin{empheq}{equation}
\varphi(0,r)=\left(2\pi\sigma_0^2\right)^{-3/2}e^{-\frac{r^2}{2\sigma_{0}^2}}.  
\end{empheq}
The Fourier transform of the initial distribution is then given by
\begin{empheq}{equation}
\varphi_{\mathbf{k}}=e^{-\frac{k^2\sigma_0^2}{2}}
\end{empheq}
and the spacetime solution $\varphi(t,r)$ written in terms of the Fourier modes in spherical coordinates becomes

\begin{empheq}{equation}
\varphi(t,r)=\frac{1}{2\pi^2r}\int_0^{\infty}k\sin(kr)e^{-\left(\frac{1}{2}k^2\sigma_0^2+i\omega(k)t\right)}dk
\end{empheq}

Continuing the Taylor expansion of the frequency to second order around $k_0=0$, using $\omega_0=\omega(k_0)=m$ and $c_g(0)=0$ gives
\begin{empheq}{align}
\omega(k)&\approx\omega_0+\frac{d\omega_0}{dk}k+\frac{1}{2}\frac{d^2\omega_0}{dk^2}k^2=m+\frac{k^2}{2m}\\
\varphi(t,r)&\approx\frac{e^{-imt}}{2\pi^2r}\int_0^{\infty}k\sin(kr)e^{-\frac{1}{2}\left(\sigma_0^2+\frac{it}{m}\right)k^2}dk\notag\\
&=\alpha(t)e^{-i\theta(r,t)}\left(2\pi\sigma_0\right)^{-3/2}e^{-\frac{r^2}{2\sigma(t)^2}}
\end{empheq}
where the spread $\sigma(t)$, amplitude $\alpha(t)$, and phase factor $\theta(r,t)$ are given by
\begin{empheq}{align}
\sigma(t)=&\sqrt{\sigma_0^2+\frac{t^2}{m^2\sigma_0^2}}\ ,\ \alpha(t)=\left(1+\frac{t^2}{m^2\sigma^4}\right)^{-3/4}\\
\theta(r,t)=&\frac{mt}{2}\left[2-\frac{r^2}{(t^2+m^2\sigma_0^4)}\right]+\frac{3}{2}\arg\left[\sigma_0^2+\frac{it}{m}\right]  
\end{empheq}
For the series expansion to be valid, the wavenumber must be much smaller than the mass of the field, $k<<m$, so the approximation is only valid for perturbations much larger that the Compton wavelength of the field, $\sigma_0>>m^{-1}$. The rate of change in the spread is given by

\begin{empheq}{equation}
c_{\sigma}=\frac{d\sigma}{dt}=\frac{1}{m\sigma_0\sqrt{1+m^2\sigma_0^4/t^2}}
\end{empheq} 
which converges towards $c_{\sigma}=(m\sigma_0)^{-1}=\sigma_k/m$ as $t\rightarrow\infty$. We note that for the region of validity of the approximation, the spread velocity is always smaller than the speed of light, $v_{\sigma}<1$, and is proportional to the mass of the scalar field and the width of the original perturbation.

\begin{figure*}[htbp!]
		\subfloat[$m=0.1$]{\includegraphics[width=0.33\linewidth]{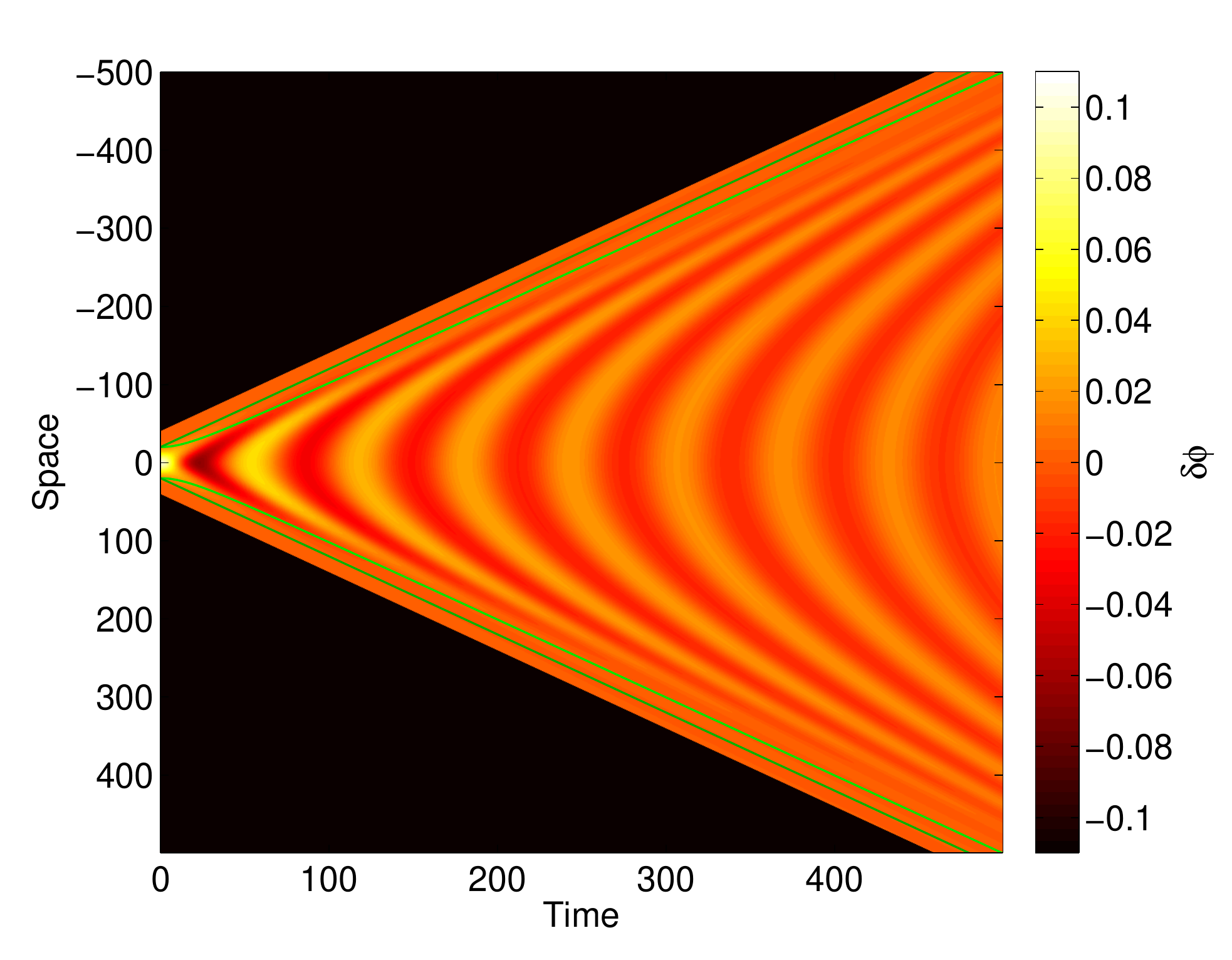}}
    \subfloat[$m=0.5$]{\includegraphics[width=0.33\linewidth]{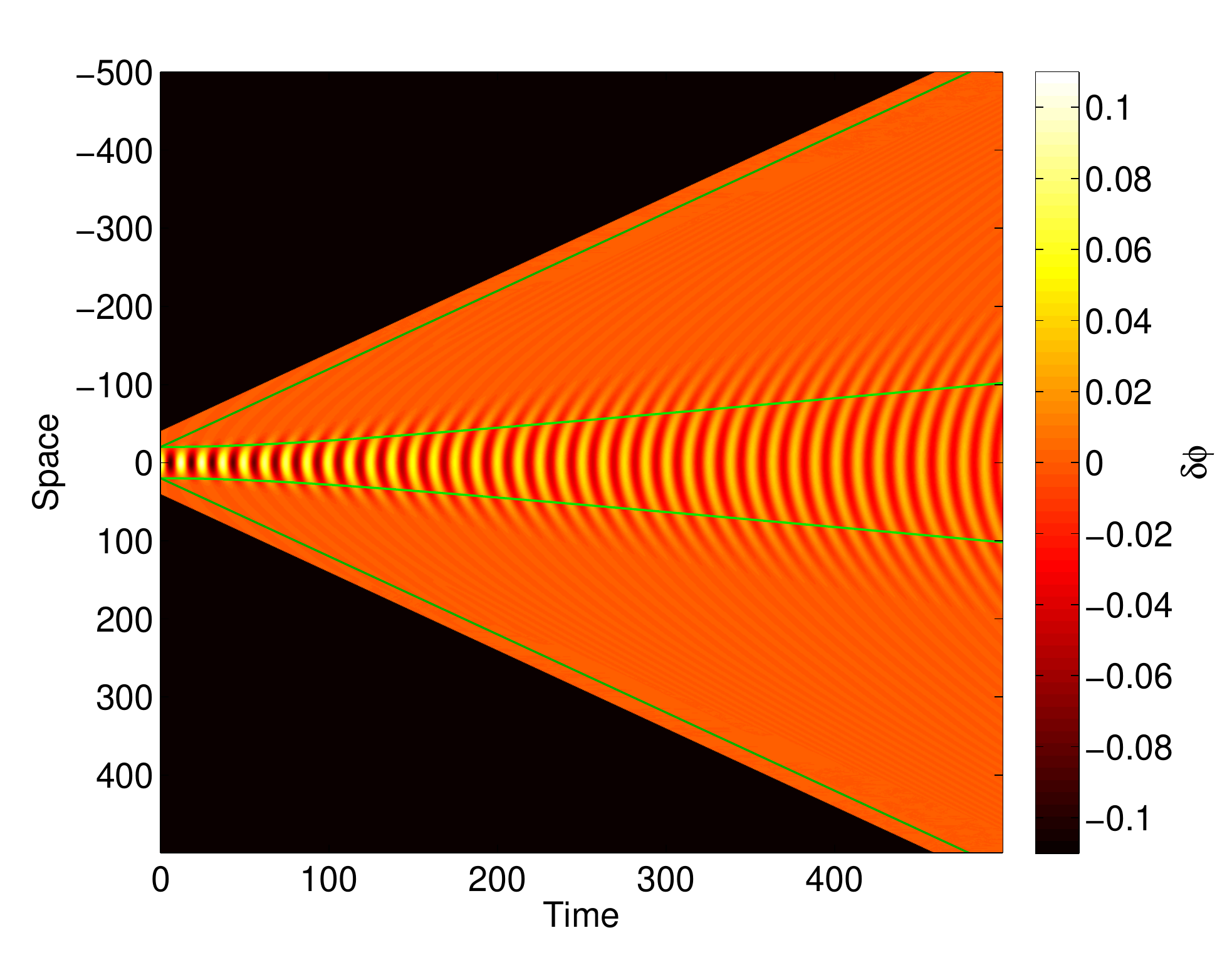}}
    \subfloat[$m=1$]{\includegraphics[width=0.33\linewidth]{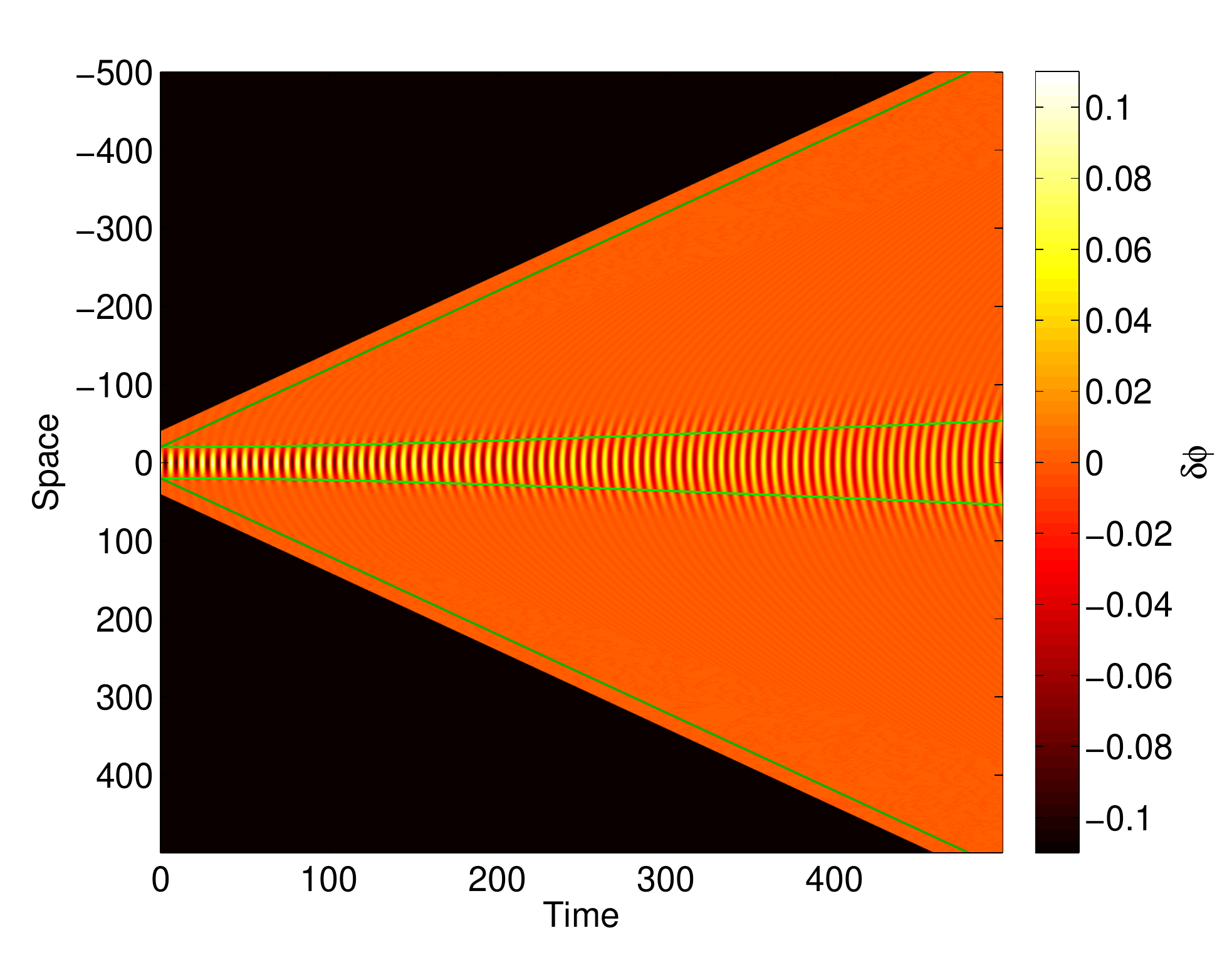}}
    \caption{The spread of a Gaussian perturbation with initial spread $\sigma_{0}=10$ for masses $m=0.1,0.5,1$. The lines indicate the paths of light rays originating from $x_c(t)=\pm(2\sigma_{0}+ct)$ (dark green, wide angle) and the spread $\pm2\sigma(t)$ (light green, small angle).}
    \label{spread_fig}            
\end{figure*}

Figure~\ref{spread_fig} shows light cone plots for the evolution of a Gaussian perturbation with initial spread $\sigma_0=10$ for three different scalar masses $m\in{0.1,0.5,1}$  as given by \eqref{Eq:pert_CEOM} with $\delta\rho=0$.  Though we have shown it here only for a Gaussian initial distribution, a similar behavior applies to more general distributions.

\section{\label{sec:ScalarWaves:NonLinReg}Non-Linear Regime}

\subsection{\label{sec:ScalarWaves:NonLin1D}Non-Linear: 1-dimensional}

We now continue by solving the full non-linear equations numerically to see how the non-linearities affect the propagation of waves that originate from a harmonically oscillating source.  Again we consider a Gaussian source $\delta\rho$ embedded in a homogeneous background $\rho_0$, where we denote the associated background value for the field by $\phi_0$. We can then write the full equations of motion in terms of the dimensionless quantities $\chi=\phi/\phi_0=1+\delta\chi$ and $\eta=\rho/\rho_0=1+\delta\eta$ as
	
	\begin{empheq}{align}
	&\nabla_{\mu}\nabla^{\mu}\chi=\frac{m^2}{n+1}\left(\eta-\chi^{-(n+1)}\right)\\
	&\left(\nabla_{\mu}\nabla^{\mu}-m^2\right)\delta\chi=\frac{m^2}{n+1}\delta\eta\ \ \ ,\ \ \ \delta\chi<<1
	\label{eq:chi_pert}
	\end{empheq} 

        The linear approximation is valid for $|\varphi|<<\phi_0$, which is equivalent to $|\delta\chi|<<1$. We can use the scaling property of the linear solution, namely that scaling the magnitude of the source corresponds to an equal scaling in the amplitude of the solution, to estimate where the linear approximation breaks down. The amplitude of the solution to equation \eqref{Eq:pert_CEOM} for $m=M=k_0=0.1$ and $\beta=1$ is roughly $\varphi\sim 0.1$, so in order to generate amplitudes, $\varphi\sim 1$, we would need $\beta M\sim 1$. In terms of the new parametrization \eqref{eq:chi_pert} using $m=0.1, n=1$ and a source $\delta\eta$ given by
	
\begin{figure*}[htbp!]
  	\captionsetup[subfigure]{labelformat=empty}
  	\subfloat{\includegraphics[width=0.33\linewidth]{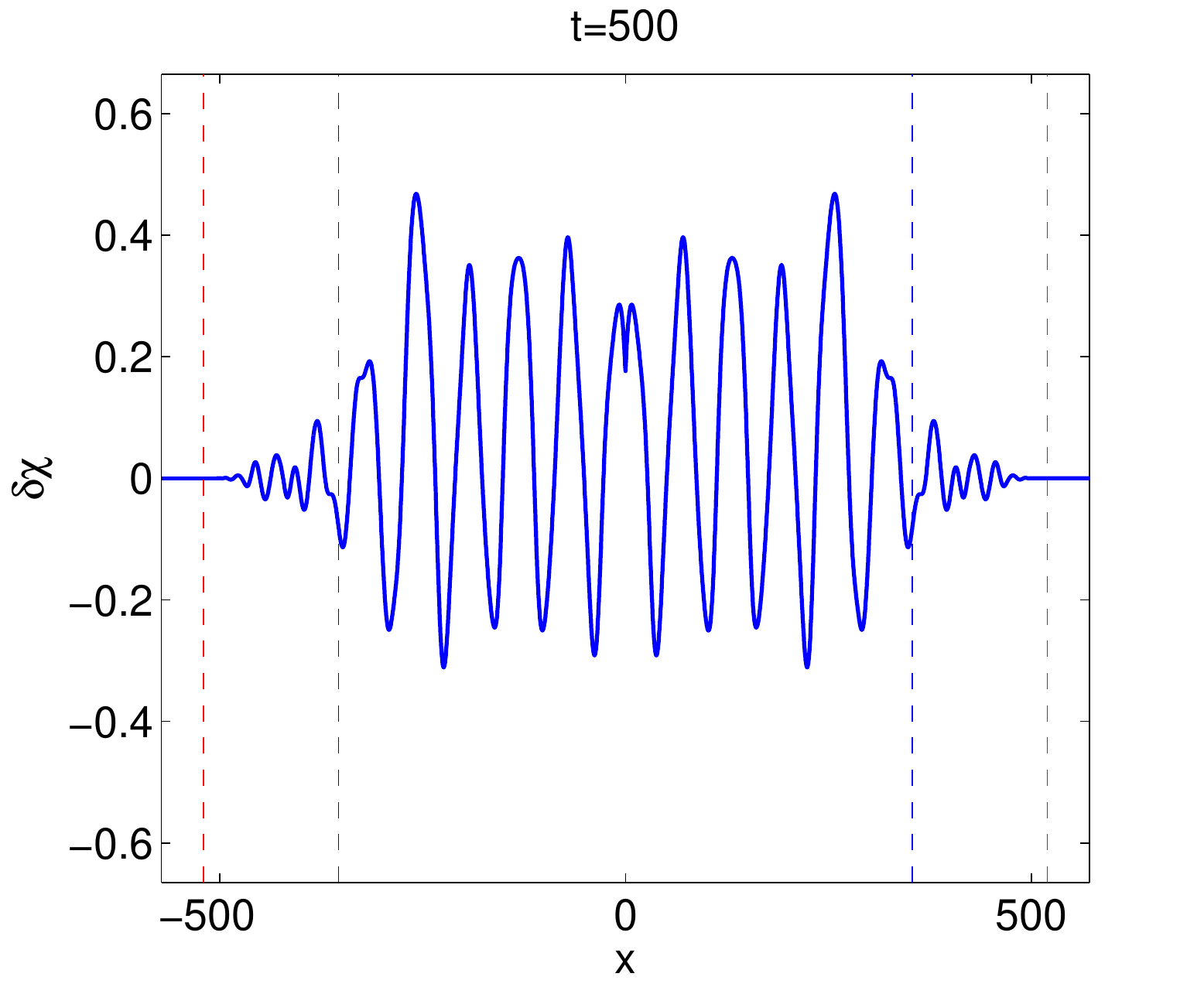}}
    \subfloat{\includegraphics[width=0.33\linewidth]{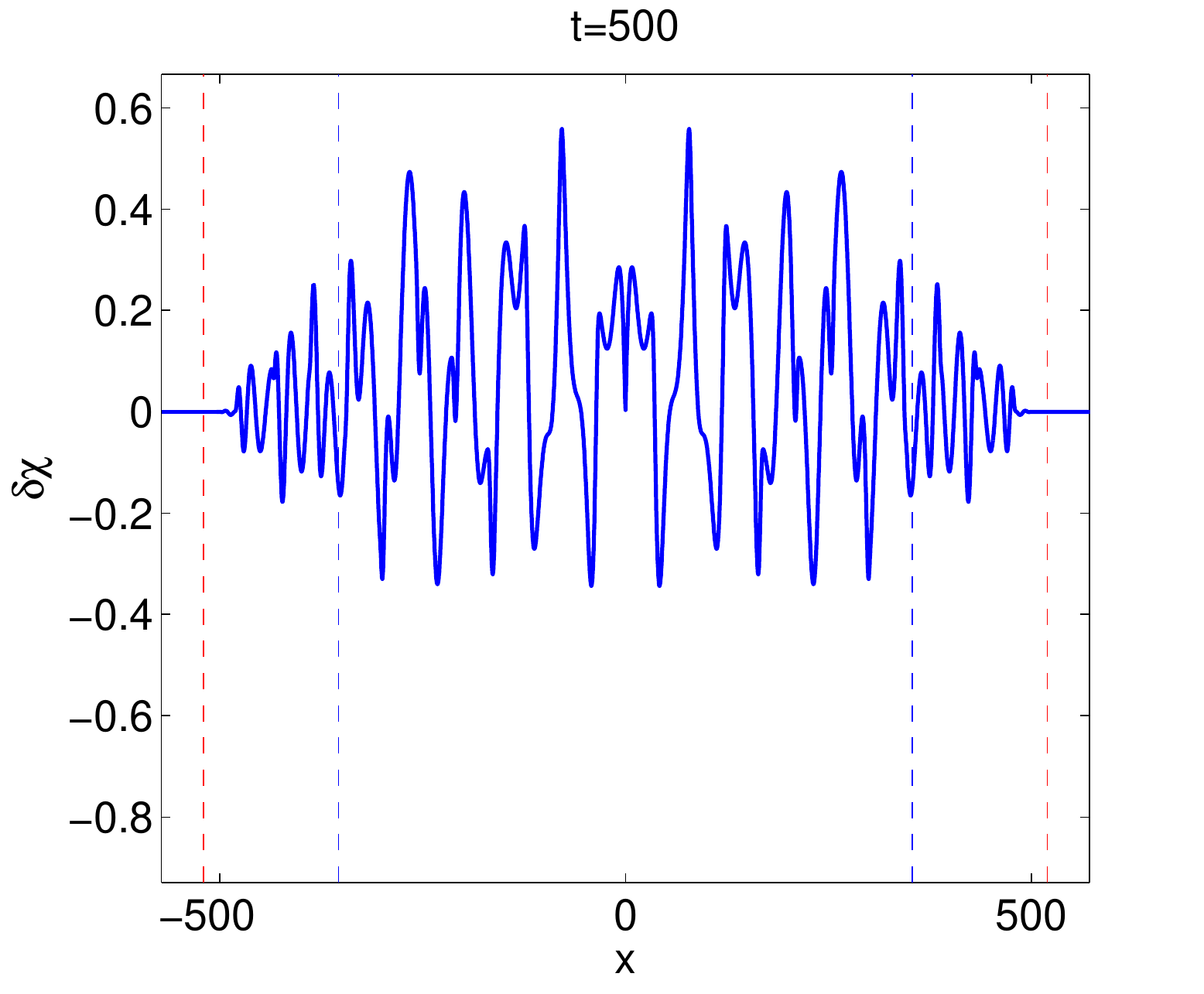}}
    \subfloat{\includegraphics[width=0.33\linewidth]{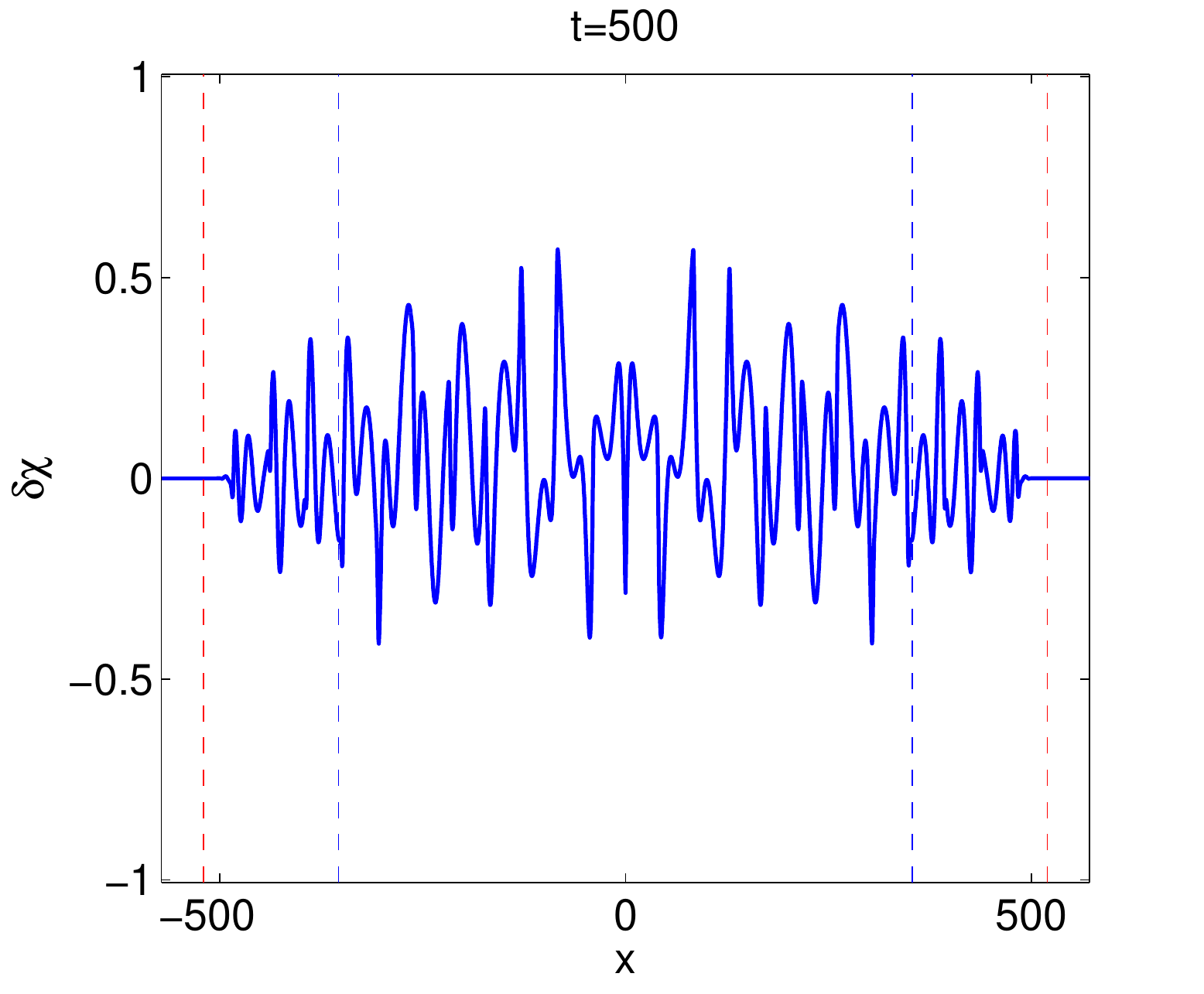}}\\
    \subfloat[(a) $\delta\eta=100$]{\includegraphics[width=0.33\linewidth]{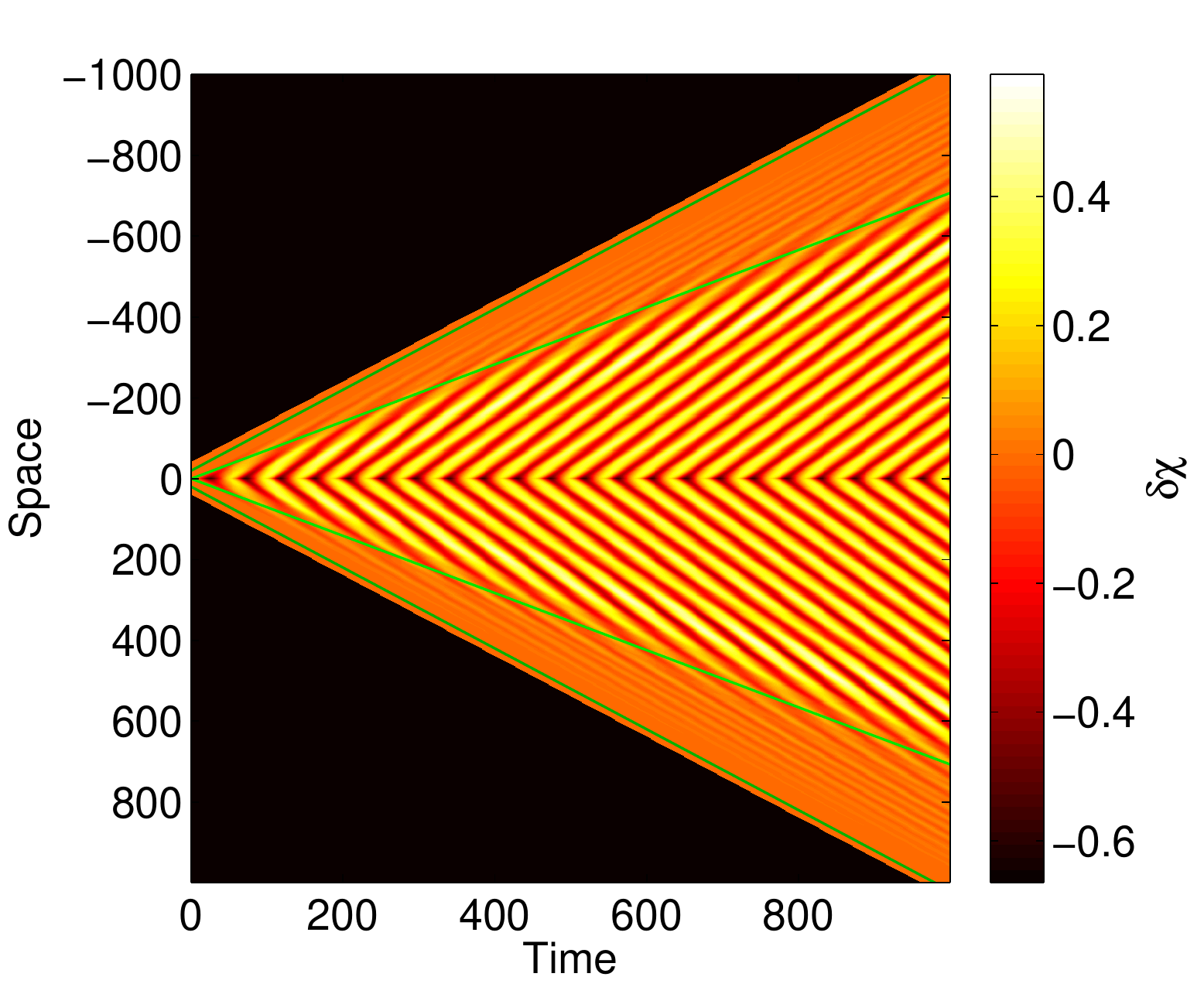}}
    \subfloat[(b) $\delta\eta=250$]{\includegraphics[width=0.33\linewidth]{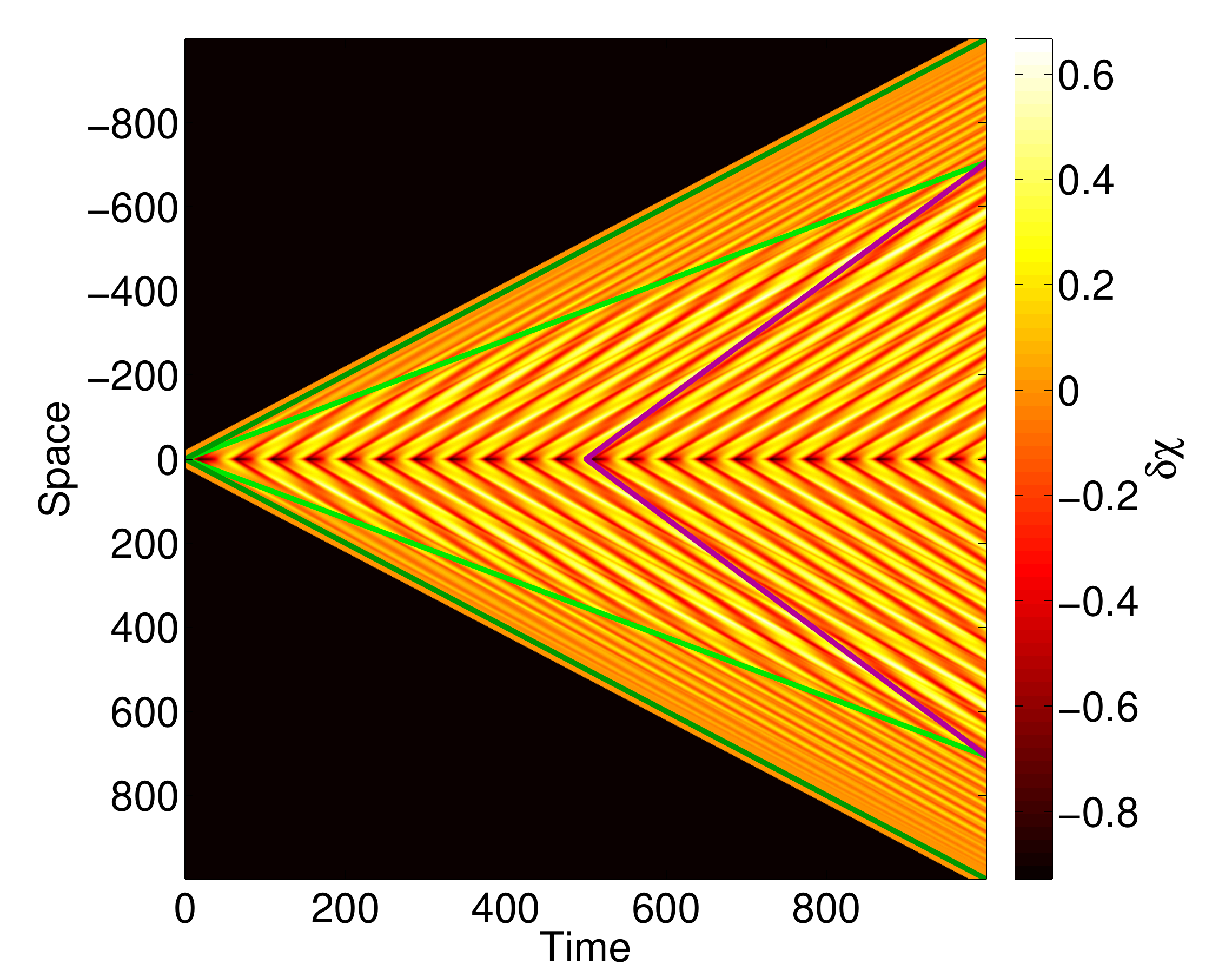}}
    \subfloat[(c) $\delta\eta=500$]{\includegraphics[width=0.33\linewidth]{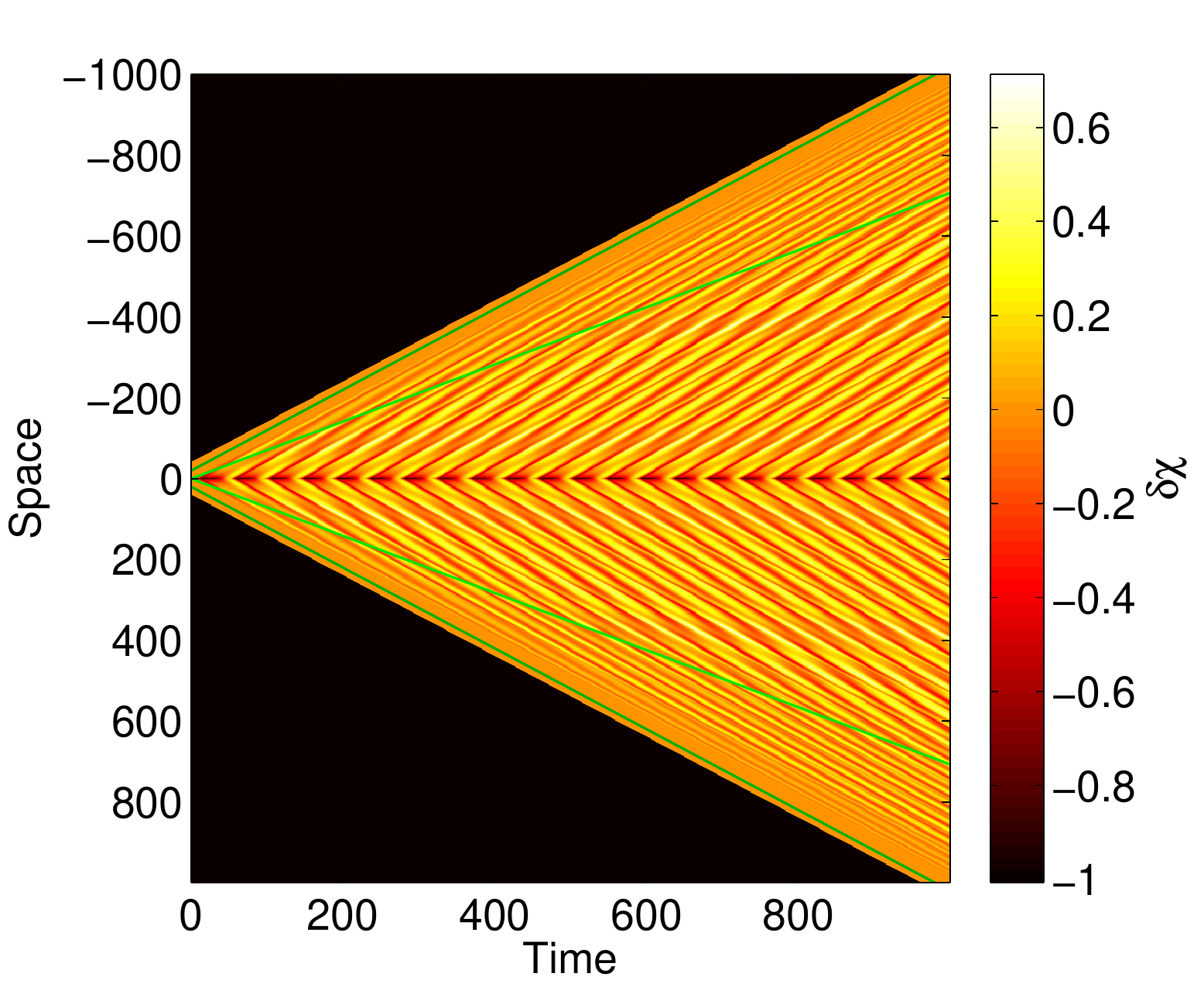}}
    \caption{Non-linear effects on the waves induced by a 1-dimensional oscillating source for three different density contrast. The upper plots show a snapshot of the profile at $t=500$, while the lower plots show the corresponding light cone plots with the light and group velocity horizons marked in light and dark green respectively. The middle plot also shows the phase velocity associated with the dominant plane wave mode of the matter source marked in purple.}
    \label{NonLinearlightcone_fig}             
\end{figure*}

	\begin{empheq}{align}
	&\delta\eta=\frac{1}{2}\delta\eta_0(x)(1-\cos(\omega_m t))\\ &\delta\eta_0(x)=Me^{-\frac{x^2}{2\sigma^2}}\ \ \ ,\ \ \ \sigma=0.1
	\end{empheq}
	
	this translates into a source amplitude of order $M\sim200$. Figure~\ref{NonLinearlightcone_fig} shows the solutions for three different density contrasts $M=100,250,500$.  
	
We clearly see that as the linear approximation breaks down, the simple oscillations in the source term no longer give a simple signal, but gives rise to complicated waveforms. We also see that the growth of the amplitudes associated with an increased density contrast is highly suppressed due to the bounds on the central oscillations discussed at the end of \ref{sec:LinearCausal}. Furthermore as the non-linearities grow, the decay of the solutions in front of the horizon associated with the group velocity becomes less evident, and the collective propagation becomes more light like. However from the light cone plots we still see a rather distinct change in the waveform at the horizon where the main line feature, corresponding to the peaks and troughs of the perturbations, decays rapidly. The motion of these peaks and troughs corresponds to the dominant plane wave mode of the signal, directly related to the main Fourier mode of the source given by equation \eqref{Eq:fourier_sol}. As can be seen from the lower middle plot in figure \ref{NonLinearlightcone_fig}, the slope of these peaks and troughs are given by the phase velocity
\begin{equation}
c^m_p=\frac{\omega_m}{k_0}=\sqrt{2}
\end{equation}
While the amplitude of the waves is greatly suppressed by the non-linearities, the main wave mode associated with the oscillation frequency of the source still travels at the group velocity.
		
\subsection{\label{sec:ScalarWaves:NonLin3D}Non-Linear: Spherically Symmetric Oscillator}

\begin{figure*}[htbp!]
		\captionsetup[subfigure]{labelformat=empty}
		\subfloat{\includegraphics[width=0.33\linewidth]{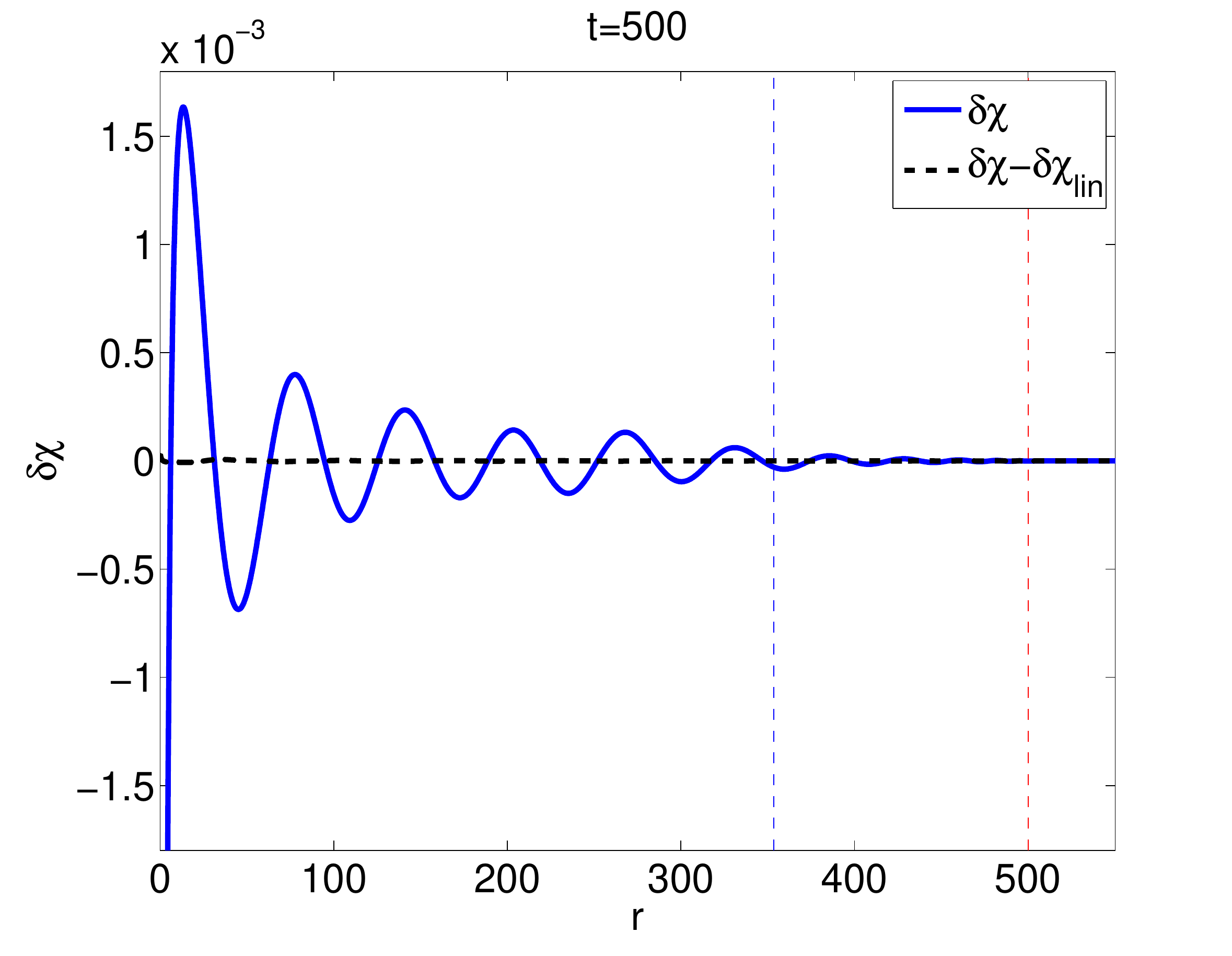}}
		\subfloat{\includegraphics[width=0.33\linewidth]{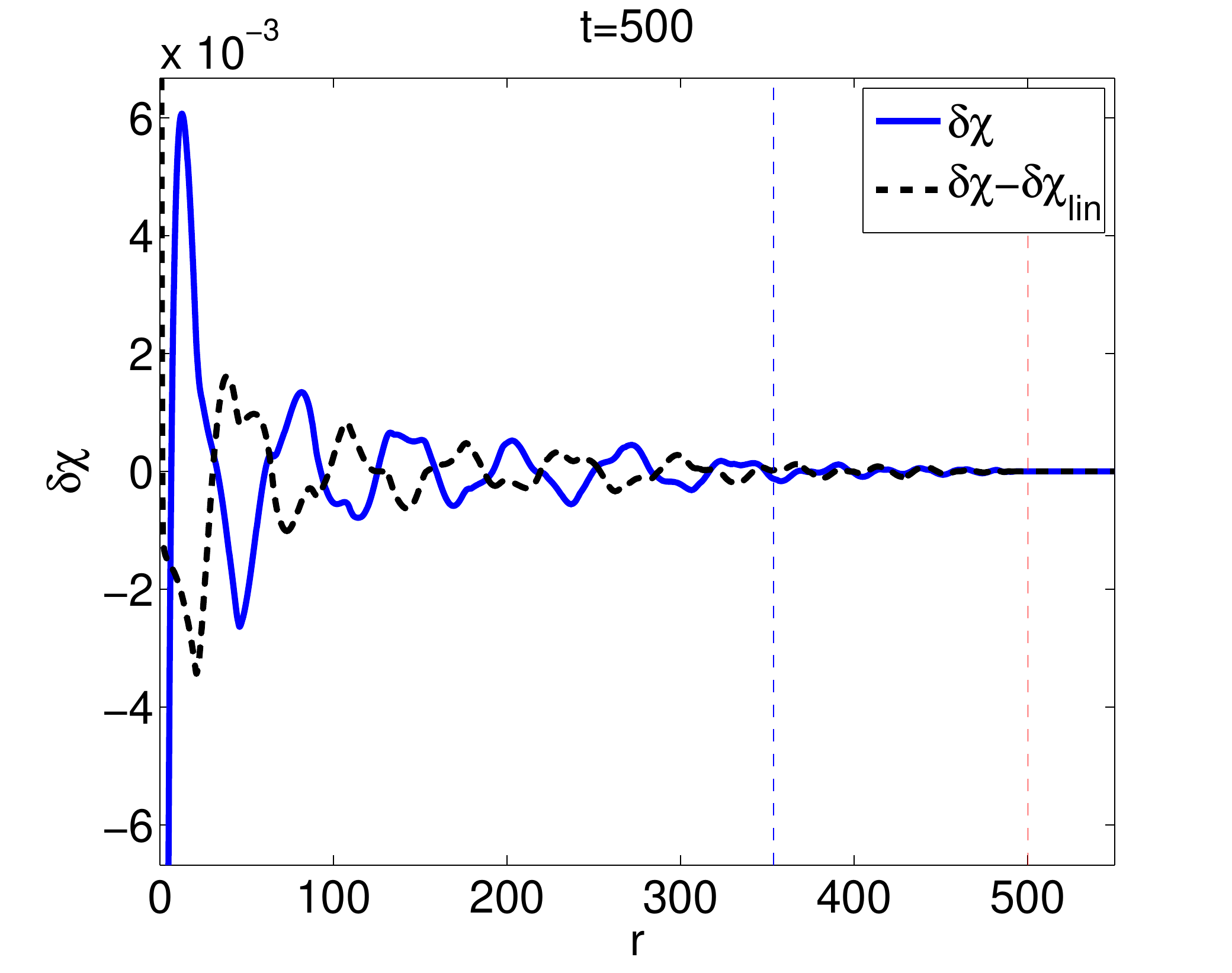}}
		\subfloat{\includegraphics[width=0.33\linewidth]{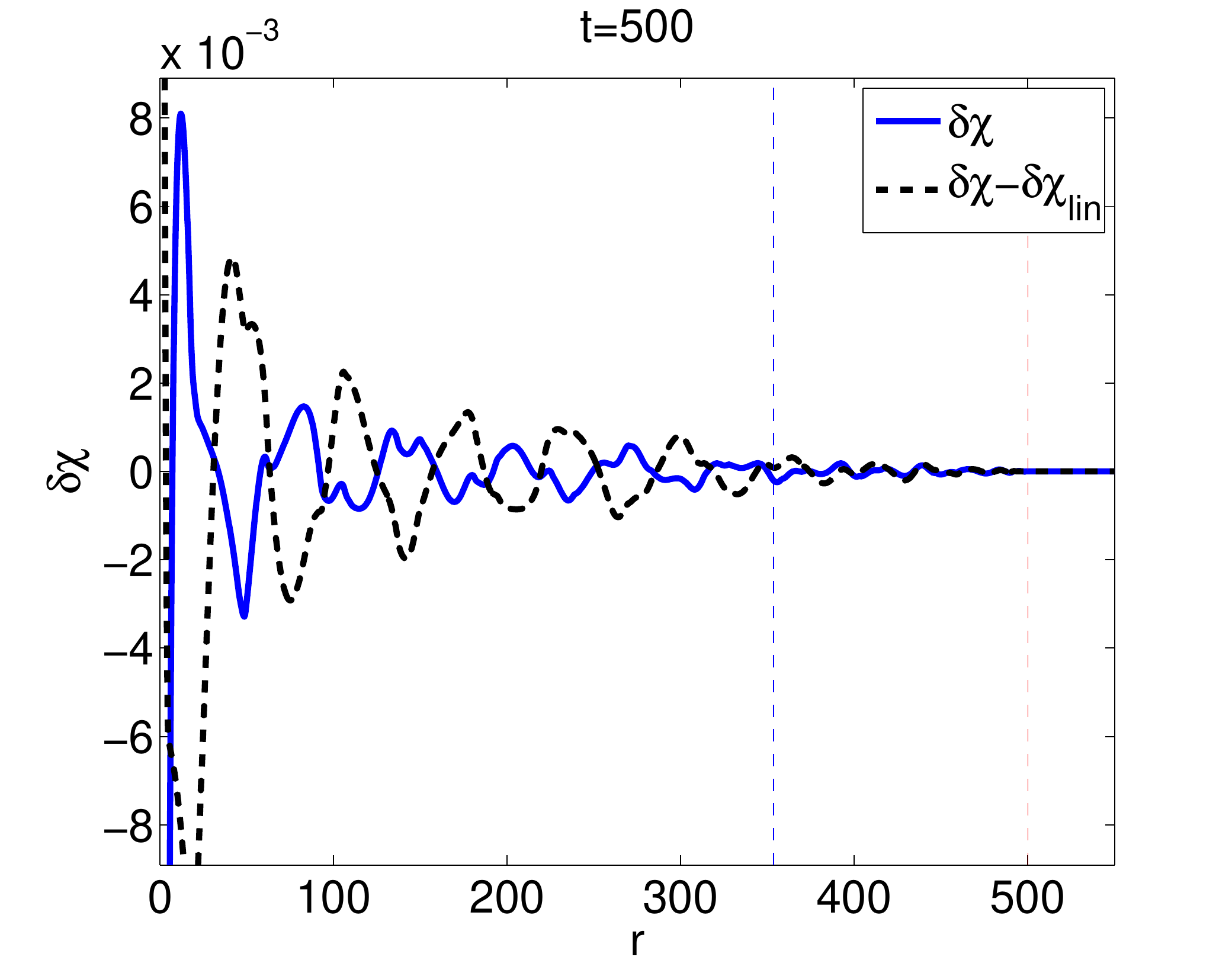}}\\
		\subfloat[(a) $M=1\cdot10^4$]{\includegraphics[width=0.33\linewidth]{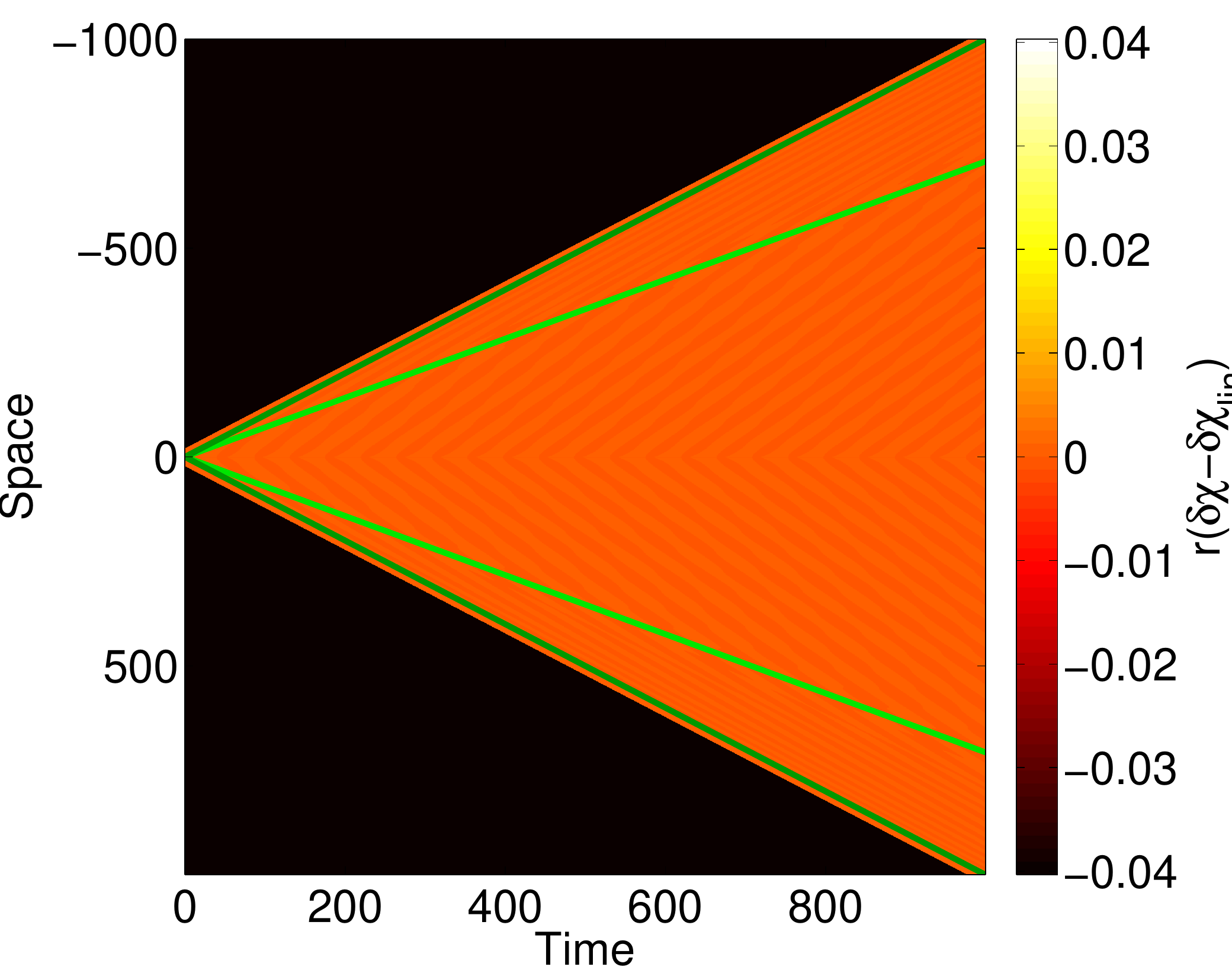}}
		\subfloat[(b) $M=5\cdot10^4$]{\includegraphics[width=0.33\linewidth]{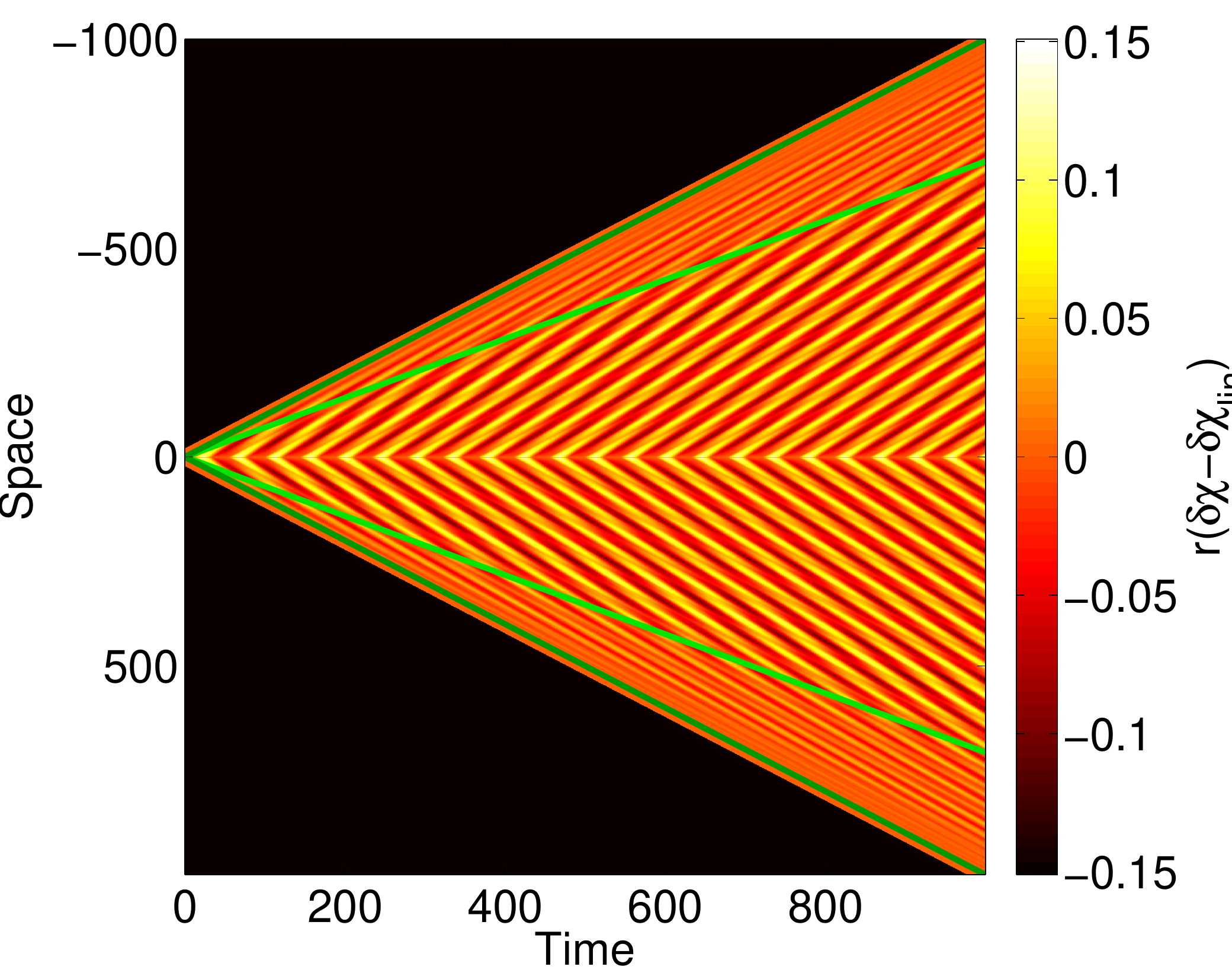}}
		\subfloat[(c) $M=10\cdot10^4$]{\includegraphics[width=0.33\linewidth]{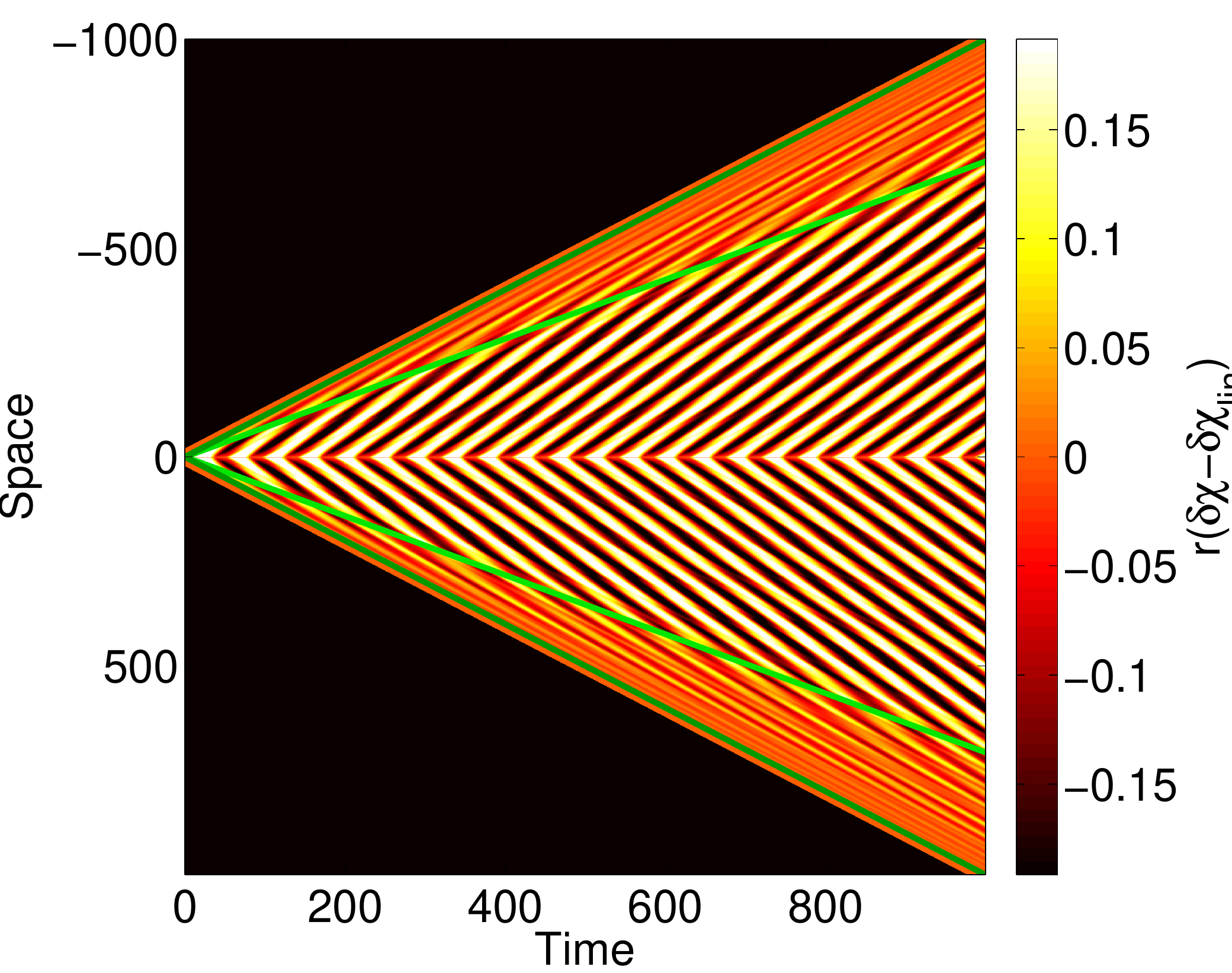}}
		\caption{Non-linear effects on the waves induced by spherically symmetric oscillating source for three different density contrast. The upper plots show a snapshot of the non-linear solution at $t=500$, together with the deviation from the linear solution. The lower plots show the corresponding light cone plots for the non-linear deviations, $r\left( \delta\chi-\delta\chi_{\text{lin}}\right) $, scaled by the radius in order to properly see the horizon. The light and group velocity horizons marked in light and dark green respectively.}
		\label{NonLinearlightconeradial_fig}             
	\end{figure*}

We now go on to consider a slightly more realistic scenario by looking at oscillations in a spherically symmetric object, where we continue to use \eqref{eq:chi_pert} to model the matter perturbation, but with $x$ replaced by a radial coordinate $r$. For a similar scenario, but with a radially pulsating source, see \cite{Silvestri:2011ch}. In the case of three spatial dimensions the perturbations in the field are similar to the 1-dimensional case except for a decay outside the source inversely proportional to the radius. This is expected as the quantity $u=r\delta\chi$ obeys exactly the same background equations as the 1-dimensional field. We also find that much larger matter perturbations are needed in order to produce sufficiently large fluctuations in the scalar field for the non-linearities to become apparent. Figure \ref{NonLinearlightconeradial_fig} shows the scalar perturbations for three different source amplitudes, $M=\left[ 1,5 ,10\right]\times 10^4 $, and only the latter two cases shows signs of non-linearities in the waveforms. In all three cases the propagation of the waves is in good agreement with that predicted by the group velocity, given by equation \eqref{eq:group_velocity}.  

\subsection{\label{sec:ScalarWaves:RealSources}Realistic Wave Sources}
The sources considered in this paper are highly contrived, as our main focus for this paper is on the propagation of scalar waves and not their generation. There are however a few realistic scenarios that can lead to the production of such scalar waves. For one they can be produced through the motion of matter sources which induces changes in the associated scalar field profile. This motion would have to be accelerated since we are dealing with a scalar field which is invariant under Lorentz transformation \cite{raey}. An observer seeing a source moving uniformly relative to him would see the associated field profile move with the source without the production of any waves as long as the field is static in the rest frame of the source. However the orbital motion of planets and stars would create disturbances in the field, as would collisions between objects. In particular binary pulsars have long been promising sources for ordinary gravitational waves as they produce gravitational quadrupole radiation. This is also the case for the additional scalar degree of freedom \cite{deRham:2012fg,Brax:2013uh}, but unlike ordinary gravity there is also the possibility for producing dipole radiation since the scalar couples to the trace of the stress-energy tensor $T^m$ rather than $T^{m}_{\mu\nu}$, thereby relieving it of the restrictions coming from conservation of energy and momentum. The potential for probing scalar-tensor theories of gravity through dipole radiation has been studied in for instance \cite{Freire:2012mg,PhysRevD.49.6892}.

The other potential source of radiation and more directly related to the toy models studied in this paper is the creation of scalar waves through time variations in the trace of the stress-energy tensor of a matter source. This can occur either through time variations in the spatial distribution of the matter source or through conversion between non-relativistic and relativistic matter which leads to changes in the trace $T_m$. Supernovae are obvious candidates for such sources where a considerable fraction of the energy density of the initial star is converted into radiation over a short period of time, thus reducing the amplitude of the source as seen by the scalar field and subsequently producing scalar waves. This process would also be more or less spherically symmetric leading to monopole radiation as in our toy model, though without the periodic oscillations. Other potential sources for monopole radiation include spherical collapse, where waves are created by the changing shape of the source distribution and variable stars which might also give rise to periodic oscillations with a more or less well defined wave number and group velocity. For constraints on screened modified gravity from monopole radiation, see for instance \cite{Upadhye:2013nfa}.

Finally, in certain models it is possible to generate cosmological scalar waves when phase transitions occur.  This was shown using N-body simulations in \cite{Llinares:2013qbh} in the context of the symmetron model.  In this particular model a phase transition takes place close to redshift zero, when a certain symmetry is broken and the scalar field suddenly changes its value from zero to one.  This sudden change gives a kick to the scalar field which is not homogeneous in space, thus producing scalar waves.  In a disformally coupled version of this model, in which the speed of sound decreases in high density regions \citep{2016A&A...585A..37H}, these waves were found to pile up inside dark matter halos. The determination of observational consequences of such effect is work in progress.

Regarding simulations, the fact that the speed of sound of scalar fields decreases with mass may be of crucial importance when working on high resolution regions.  The reason for this is that in theories that include screening mechanisms, scalar fields typically become massive when the density increases (i.e. in the centre of the dark matter halos).  While quasi-static simulations are likely to give inaccurate results under these conditions, it is also important to keep in mind that in these high density regions, the scalar fields will be screened and thus there will be no measurable modified gravity effects in the quasi-static regime.  Experiments similar to those presented in \cite{2014PhRvD..89h4023L}, but with higher resolution, are necessary to fully understand if the screening will cancel the dynamical effects associated with the low speed of sound.  A failure of the quasi-static approximation inside dark matter halos should not be considered as a problem, but as an opportunity to find new observables that might provide a unique signature of modify gravity. The modelling of the collision between two galaxies provide one scenario where dynamical effects might be important. In such a collision, the matter distribution will suddenly change its trajectory, leading to a displacement between the perturbations in the matter distribution and the scalar field. If the speed of sound is small, a part of the original perturbation in the scalar field is expected to continue along the original trajectory of the galaxies undistorted up to a phase, see Figure \ref{spread_fig}. Such effects can only be confirmed through high resolution N-body simulations.
	
\section{\label{sec:SpeedofSound}On The Effective Speed of Sound}

Finally we briefly comment on the relation between the group velocity $c_g$, describing the propagation of linear waves in the scalar field, and the speed of sound $c_s$ appearing in the perfect fluid description of the scalar field in cosmological perturbation theory. Einsteins equations are sourced by the energy-momentum tensor $T_{\mu\nu}$, which for a canonical scalar field takes the form
\begin{empheq}{equation}
T_{\mu\nu}=\partial_{\mu}\phi\partial_{\nu}\phi-g_{\mu\nu}\big(\partial^{\mu}\phi\partial_{\mu}\phi+V(\phi)\big).  
\end{empheq}
This can be written in perfect fluid form by associating a rest frame energy density $\rho_{\phi}$ and pressure $p_{\phi}$, and a fluid four velocity $U_{\mu}$ with the scalar field

\begin{empheq}{align}
&U_{\mu}\equiv\frac{\partial_{\mu}\phi}{|\partial^{\alpha}\phi\partial_{\alpha}\phi|^{1/2}}\\
&\rho_{\phi}\equiv-\frac{1}{2}\partial^{\alpha}\phi\partial_{\alpha}\phi+V(\phi)\\
&p_{\phi}\equiv-\frac{1}{2}\partial^{\alpha}\phi\partial_{\alpha}\phi-V(\phi)
\end{empheq}
so that $T_{\mu\nu}$ takes the form

\begin{empheq}{equation}
T_{\mu\nu}=(\rho_{\phi}+p_{\phi})U_{\mu}U_{\nu}+p_{\phi}g_{\mu\nu}
\end{empheq}

By considering perturbations to the energy density and pressure, $\delta\rho_{\phi}$ and $\delta p_{\phi}$, to the cosmological FLRW background evolution, one finds that the clustering of the scalar fluid is determined by an effective speed of sound $c_s$ \cite{Garriga:1999vw}

\begin{empheq}{equation}
\label{Eq:c_seff}
c_s^2=\frac{\partial_X p_{\phi}}{\partial_{X}\rho_{\phi}}\ \ \ ,\ \ \ X=\frac{1}{2}\partial_{\mu}\phi\partial^{\mu}\phi
\end{empheq}

For a scalar field with a canonical kinetic term the effective speed of sound is equal to the speed of light, preventing the field from clustering inside the horizon \cite{Bertacca:2010ct}. Since the speed of sound determines how small perturbations in a fluid propagates, this is sometimes taken to mean that perturbations in a canonical scalar field always propagate at the speed of light. However the speed of sound as defined in \eqref{Eq:c_seff} refers to the propagation of perturbations in the energy density and pressure as seen from the fluid rest frame, defined as the frame where spatial variations in the scalar field vanish $\partial_i\phi=0$. This is clearly not the case in the rest frame of the matter source which is what we are considering here.

The perfect fluid description also puts constraints on the allowed dynamics of the scalar field. In particular the gradients of the field are constrained by the requirement that $U^{\mu}$ be a proper four velocity \cite{Sawicki:2012re}. 

\begin{empheq}{equation}
U_{\mu}U^{\mu}=-1\ \ \ ,\ \ \ U_0>0
\end{empheq}
This implies that the derivatives of the scalar field must be non-vanishing $|\partial_{\mu}\phi|>0$, and that the time derivative of the field $\dot{\phi}$ cannot change sign. None of these conditions are satisfied for the simple static background approximation we have employed in studying the propagation of perturbations in the field. In addition, as already mentioned the speed of sound is defined with respect to the rest frame of the scalar fluid, which is a natural reference frame when the only degrees of freedom are the metric and the scalar field, but not when considering scalar waves induced by a matter source, where the natural frame is the matter rest frame. In this frame the scalar fluid has a non-vanishing four velocity and hence the flow of energy and momentum is different from that of the scalar field reference frame. For example the energy density $\tilde{\rho}_{\phi}$ as seen by an observer moving with a four velocity $U_{\mu}$ relative to the scalar fluid is given by

\begin{empheq}{equation}
\tilde{\rho}_{\phi}=T_{\mu\nu}U^{\mu}U^{\nu}
\end{empheq}
which means that in the matter rest frame energy propagates as a combination of sound waves with respect to the fluid rest frame and through the flow of the fluid itself. In fact in \cite{Hu:2000ke}, they consider a classical scalar field coupled to gravity and find an effective speed of sound

\begin{empheq}{equation}
c_s^2=\frac{\delta p}{\delta\rho}\approx\frac{k^2}{4m_{\phi}^2}
\end{empheq}
This is similar to the limiting spread velocity $c_{\sigma}$ for a Gaussian distribution discussed in section~\ref{sec:LinearCausal}, since the spread of a Gaussian in real space is the inverse of the spread in Fourier space, $\sigma_0=\sigma_k^{-1}$. Much more detailed accounts regarding the correspondence between a perfect fluid and a scalar field can be found in the literature (see f.ex \cite{Bilic:2008yr}), we simply note that the speed of sound appearing in the fluid description and the group velocity associated with the propagation of scalar waves are different concepts and should not be confused.         

\section{\label{sec:Summary}Summary and Conclusions}

We have looked at the propagation of scalar waves induced by matter sources in the context of modified theories of gravity which include screening mechanisms.  The usual approach when studying these theories in the non-linear regime of cosmological perturbations is based on the assumption that scalar waves travel at the speed of light.  Within the context of standard gravity, it has been shown that in this case the speed can be approximated by infinity without loss of accuracy in the estimation of observables \citep{2011PhRvD..83l3505C, 2012PhRvD..85f3512G}.  However, there is no study within modified gravity that supports this idea.  Here we review the concepts of phase and group velocity in the context of modified gravity and find that mass terms and non-linearities in the equations of motion can lead to propagation and dispersion velocities significantly different from the speed of light, contrary to usual beliefs.

The origin of the assumption of light like propagation for the additional scalar degree of freedom seems to originate from the speed of sound being equal to the speed of light in the perfect fluid approximation of the scalar field often used in cosmological perturbation theory. We point out that the necessary conditions for this approximation to be valid breaks down when considering scalar waves induced by matter sources, and that the propagation of these waves is best described by the group velocity.  As the group velocity is the one associated with the propagation of signals, a reduction of its value will have direct impact on the behavior of these theories.

On cosmological scales, the background matter density and perturbations are small and slowly varying.  For the class of models studied in this paper, the scalar becomes very light and the large scale perturbations in the field will be linear for reasonable choices of model parameters. We therefore expect the assumption of light like propagation to be valid in this regime. We stress that this expectation only applies to models whose equation of motion reduces to the Klein Gordon equation (\ref{Eq:pert_CEOM}) in the linear regime, and does not extend to, for instance,  models with non-canonical kinetic terms. On the other hand, the internal dynamics of galaxies and satellites submerged in large dark matter halos could be affected by the fact that the group velocity is small.  In these cases, it will be necessary to take into account the fact that different parts of a galaxy will see changes in the environment at different times (i.e. a full non-static analysis should be made). Full 3D high-resolution simulations are required to measure the impact of this phenomenon in more realistic scenarios.

Furthermore we consider a possible link between gravitational clustering of canonical scalar fields and the dispersion rate of perturbations in the field. We find that this rate is closely related to the speed of sound related to the time averaged energy density and pressure of spatially localized scalar perturbations. In order to establish whether this spread velocity is directly related to the jeans length of the field requires the inclusion of gravitational effects on the perturbations which was beyond the scope of this paper. 

Finally, we found that non-linear terms in the equations of motion for the scalar field induce small scale oscillations superimposed on the main waves generated by the source.  Furthermore these non-linearities can reduce the amplitude of the oscillations compared to the linear case. Despite this we find that the propagation speed of the main waves is still described by the group velocity. This constitutes a characteristic signature of modified gravity which could be potentially observed with future gravitational waves experiments (e.g. LISA).  It is important to note that this observable is not degenerate with other effects. This differentiates it from other observables such as the power spectrum of cosmological density perturbations, for which the modified gravity signals are degenerated with the effects of baryons and neutrinos.  Targeted studies on specific astrophysical objects should be made to predict the amplitude of these waves under realistic situations.

Our study is focused on a particular set of models where the matter coupling is conformal, but considering that models with more exotic matter couplings usually lead to considerable reductions in the speed of sound already in the perfect fluid approximation, the results are expected to be more general. More studies are required before having a complete picture of the problem and the validity of small scale cosmological simulations. While the results found in this paper seem to constitute a problem for the community of simulations, we should look at the bright side and keep in mind that this new effect could lead to a completely new set of observables that can be used to test these theories for gravity.

\section{Acknowledgments} 
DFM acknowledges funding from the Research Council of Norway and the use of computing facilities of the NOTUR cluster HEXAGON. CLL acknowledges support from the Research Council of Norway through grant 216756 and from STFC consolidated grant ST/L00075X/1. J\O L acknowledges support from the Centre for Dark Matter Research
(DAMARA) at the Department of Physics and Technology, University of Bergen, funded by the Bergen Research Foundation, the University of Bergen, and the Norwegian Research Council.

%

\end{document}